\documentclass[prb,aps,groupedaddress,showpacs,twocolumn]{revtex4}

\usepackage{graphicx}
\def\figwidth{0.9 \columnwidth}
\def\largewidth{\columnwidth}

\begin{document}
\title{Interacting fermions in self-similar potentials}
\author{Julien Vidal}
\email{vidal@gps.jussieu.fr}
\affiliation{Groupe de Physique des
Solides, CNRS UMR 7588, Universit\'{e}s Pierre et Marie Curie
Paris 6 et Denis Diderot Paris 7, 2 place Jussieu, 75251 Paris
Cedex 05 France}
\author{Dominique Mouhanna}
\email{mouhanna@lpthe.jussieu.fr} \affiliation{Laboratoire de
Physique Th\'eorique et Hautes \'Energies, CNRS UMR 7589,
Universit\'e Pierre et Marie Curie Paris 6, 4 place Jussieu,
75252 Paris Cedex 05 France}
\author{Thierry  Giamarchi}
\email{giam@lps.u-psud.fr} \affiliation{Laboratoire de Physique des
Solides, CNRS UMR 8502,
Universit\'e Paris Sud, B\^atiment  510, 91405 Orsay France}


\begin{abstract}
We consider interacting spinless fermions in one dimension embedded in
self-similar quasiperiodic potentials. We examine generalizations of
the Fibonacci potential known as precious mean potentials. Using a bosonization
technique and a renormalization group analysis, we study the
low-energy physics of the system. We show that it undergoes a metal-insulator
transition for any filling factor, with a critical interaction that
strongly depends on the position of the Fermi level in the Fourier
spectrum of the potential. For some positions of the Fermi level the 
metal-insulator
transition occurs at the non interacting point. The repulsive side is 
an insulator
with a gapped spectrum whereas in the attractive side the spectrum is
gapless and the properties of the system are described by a
Luttinger liquid. We compute the transport properties and give the 
characteristic
exponents associated to the frequency and temperature dependence of the
conductivity.
\end{abstract}

\pacs{61.44.Br, 71.10.-w, 71.30.+h}

\maketitle


\section{Introduction}

The electronic properties of
quasicrystals\cite{shechtman_quasi_discovery} have revealed the importance
of the non crystalline order at the atomic level. Indeed, the
conductivity $\sigma$ of
these metallic alloys displays a unusual behavior since it
increases when either temperature or disorder increases. It is also
surprisingly low compared to that of the metals that composed them.
    From a theoretical point of view, the influence of quasiperiodicity
on the spectral and dynamical properties of electron systems has
been the subject of many
studies\cite{Sire_aussois,schulzbaldes_aperiodic_transport,piechon_dif 
fusion_fibo,%
sire_quasi_gaps,ketzmerick_quasi_wavepacket,roche_review_quasi}.
For independent electrons systems, it has been shown that the
eigenstates, which are
neither localized nor extended but critical (algebraic decay),
are responsible
of  an anomalous quantum diffusion in any dimension. Concerning the
nature of the spectrum, it depends on the dimensionality but also exhibits
specific characteristics of the quasiperiodicity.
More precisely, in one dimension, the spectrum of quasiperiodic systems,
such as the
Fibonacci or the Harper chain, is made up of an infinite number of
zero width bands
(singular continuous) whereas in higher dimensions, it can be either absolutely
continuous (band-like), singular continuous, or any mixture.
These features are a direct consequence of the long-range order
present in these
structures despite the lack of periodicity.
This absence of translational invariance makes any analytical
approach difficult and one
must often have recourse to numerical diagonalization, except in a perturbative
framework\cite{sire_quasi_gaps,piechon_diffusion_fibo}.

Given the complexity of the independent electron problem, the
influence of a quasiperiodic modulation on an
interacting system is very difficult to tackle. Attempts
to solve this problem have been mostly confined to mean field
solutions\cite{hiramoto_hartreefock_quasi} or numerical diagonalizations
\cite{chaves_meso_transport,chaves_harper_meanfield,eilmes_mit_twoparticles}.
We have recently proposed \cite{vidal_quasi_interactions_short} a
different route,
already used with success for periodic
\cite{giamarchi_umklapp_1d,giamarchi_mott_shortrev} and
disordered systems\cite{apel_spinless,giamarchi_loc}. The main idea
of this method is to
first solve the {\it periodic} system in presence of interactions~; this is
relatively easy, either in the one-dimensional case  for which
technique to treat interactions
exists\cite{solyom_revue_1d,emery_revue_1d,schulz_houches_revue,voit_b 
osonization_revue}, or
even in higher dimensions through approximate (Fermi liquid)
solutions. In a second step, we study the effect of a perturbative
quasiperiodic potential via a renormalization group approach.
Several types of quasiperiodic potentials can in principle
be studied by this approach but the most interesting effects come from
quasiperiodic potentials which have a non trivial Fourier spectrum.
Indeed other potentials such as the Harper
model\cite{kolomeisky_harper_perturbatif,sen_quasi_dimerization,sen_qu 
asi_renormalization}
who have only a single harmonic in their Fourier spectrum are
perturbatively equivalent to periodic
systems
\cite{giamarchi_umklapp_1d}.
We have used our RG approach to treat interacting spinless fermions
in the presence of a Fibonacci
potential\cite{vidal_quasi_interactions_short,vidal_quasiinter_mbx}.
We have shown that the existence of arbitrarily small peaks in the
Fourier spectrum (opening arbitrarily small gaps at first order in
perturbation) leads
to a vanishing critical interaction below which the system is conducting.
This novel metal-insulator transition (MIT) has very different
characteristics from those observed in periodic and disordered
systems for which a finite attractive
interaction is required.
These predictions have been successfully confirmed by numerical
calculations\cite{Hida_quasi_spinless_DMRG,Hida_precious}.
Similar renormalization techniques have been also used in a variety of cases
\cite{mastropietro_quasi_smalldenominators,sen_quasi_dimerization,sen_ 
quasi_renormalization,Hida_quasi_spinfull}.
Even if some of these properties are specific to  one-dimensional
potentials, these results should provide a first step toward the
understanding of higher
dimensional interacting system in quasiperiodic structures.

In the present paper, we extend this study to quasiperiodic
potentials that generalize the Fibonacci potential.
We show that the critical properties obtained in the Fibonacci
case\cite{vidal_quasi_interactions_short} are generic of other
self-similar systems.
Our results are in agreement with the recent numerical results
obtained on precious mean
potentials\cite{Hida_precious}. The paper is  organized as follows~:
in Section~\ref{The_Model}, we
present the model on the
lattice and derive its continuous version for any  potential using a
bosonization technique. We detail the
renormalization group treatment of the bosonized model and
the computation of the flow equations for the coupling constants. In
Section~\ref{Critical_properties}, we recall the results for the well-known
Mott transition
(periodic case) and we  describe the physics of the disordered
case for which a
different kind of MIT occurs. We then discuss the most interesting
situation~: the
quasiperiodic case. We explain why the non trivial self-similar
Fourier spectrum induces a
MIT whose characteristics are intermediate between the periodic and
the disordered potentials. The physical consequences are discussed in
the Section~\ref{Transport} with a special emphasis on the transport
properties. We  also discuss
the question of the strong coupling regime. Conclusions can be found
in Section~\ref{sec:conclusions}
and some technical details are given in the appendices.

\section{Description of the Model and Renormalization}
\label{The_Model}

\subsection{The Model}

We consider a system of interacting spinless fermions in a
one-dimensional lattice of linear size $La$
($a=1$ being the lattice spacing) described by the following Hamiltonian~:
%
%
\begin{eqnarray}
H&=&-t \sum_{i} c^\dagger_{i+1}\,c_{i}+ c^\dagger_{i}\,c_{i+1}+ \nonumber\\
&&V \sum_{i}  n_i\,n_{i+1} +
\sum_{i} W_i\,n_i \label{eq:hamiltonian} \\
&=& H_t+H_V+H_W
\end{eqnarray}
%
%
where $c^\dagger_{i}$ (resp. $c_{i}$) denotes the creation (resp.
annihila\-tion) fermion operator, $n_i=c^\dagger_{i}\,c_{i}$
represents the fermion density on site $i$. In
(\ref{eq:hamiltonian}), $t$ represents the hopping integral
between sites and $V$ controls the strength  of the interaction
between  nearest-neighbor particles. In addition, the fermions are
embedded in  an on-site (diagonal) potential $W$. In the following,
we consider three main categories for $W$~: $(i)$ a simple
periodic potential of the form $W_i = \lambda \cos(Q i)$; $(ii)$ a
random potential uncorrelated from site to site; $(iii)$ a
quasiperiodic potential whose study is the aim of this
paper. In this latter case, we will focus on the general class of precious mean
potentials described in  Appendix \ref{ap:precious}.

In order to treat the interactions in (\ref{eq:hamiltonian}), it
is convenient to write the fermion operators in term
of boson ones. This bosonization
technique\cite{solyom_revue_1d,emery_revue_1d,schulz_houches_revue,voi 
t_bosonization_revue} provides
a good description of the low-energy physics of the Hamiltonian
$H_0(t,V)=H_t+H_V$.
For completeness and to fix the notations we give a brief summary of
this method in
Appendix~\ref{ap:bosonization}. Within this framework,
$H_0(t,V)$ can be rewritten~:
%
%
\begin{equation}
\label{eq:ham_boso_texte}
H_0(t,V)={1 \over 2 \pi} \int dx  \: (uK) (\pi \Pi)^2 +
\left({u \over K}\right)
(\partial_x \phi)^2
\mbox{,}
\end{equation}
%
%
where $\phi$ and $\Pi$ are conjugate bosonic fields respec\-tively related to
the long wavelength part of
the density and the current
(see Appendix~\ref{ap:bosonization}). All interaction effects can
be absorbed in the so-called Luttinger liquid  parameters~: $u$, the velocity
of the charge excitations, and $K$ which controls the behaviour of
the various correlation functions
(see below). For  $V \ll t$, analytic expressions of these parameters
can be obtained (see (\ref{eq:pertlut2}-\ref{eq:pertlut})).
However,  the bosonic representation is in fact quite general and the
expression
(\ref{eq:ham_boso_texte}) provides the effective Hamiltonian
describing the low-energy physics of any
one-dimensional interacting spinless fermionic system
\cite{haldane_xxzchain,haldane_bosonisation}.

Concerning the coupling to the lattice potential $W$, one has, in the
continuum limit
(see Appendix~\ref{ap:bosonization})~:
\begin{eqnarray}
H_{W}&=& \int  dx \: W(x) \,\rho(x) \\
&=&\int dx \: W(x) \times  \nonumber\\
&& \left(-{1 \over \pi} \partial_x
\phi(x) + {1 \over \pi \alpha} \cos(2\phi(x)-2k_F x) \right)
\label{eq:hwfinal}
\mbox{.}
\end{eqnarray}

The various physical observables can be expressed in terms of the
boson fields. For example,
the correlation function of the $2k_F$ part of the density is~:
\begin{eqnarray}
R(x,\tau,x',\tau')&=& \left\langle T_\tau \rho_{2k_F}(x,\tau) \:
\rho_{2k_F}(x',\tau')
\right\rangle \\ &\sim&\left\langle T_\tau e^{ i 2\phi(x,\tau)}
e^{-i 2\phi(x',\tau')}\right\rangle
\mbox{,}
\label{eq:correlationboso}
\end{eqnarray}
where $T_{\tau}$ is the time-ordering operator for the imaginary time $\tau$.
In absence of the perturbation $H_W$, one has~:
\begin{equation}\label{scalingfree}
R(x,\tau,x',\tau') \sim \left({\alpha \over |{\bf r
-r'}|}\right)^{2K} \hbox{for}\
\ |{\bf r -r'}|\gg \alpha
\mbox{,}
\label{decaying}
\end{equation}
where ${\bf r}=(x,\tau)$.

\subsection{Renormalization Group Analyzis}

\begin{figure*}
\centerline{\includegraphics[width=\figwidth]{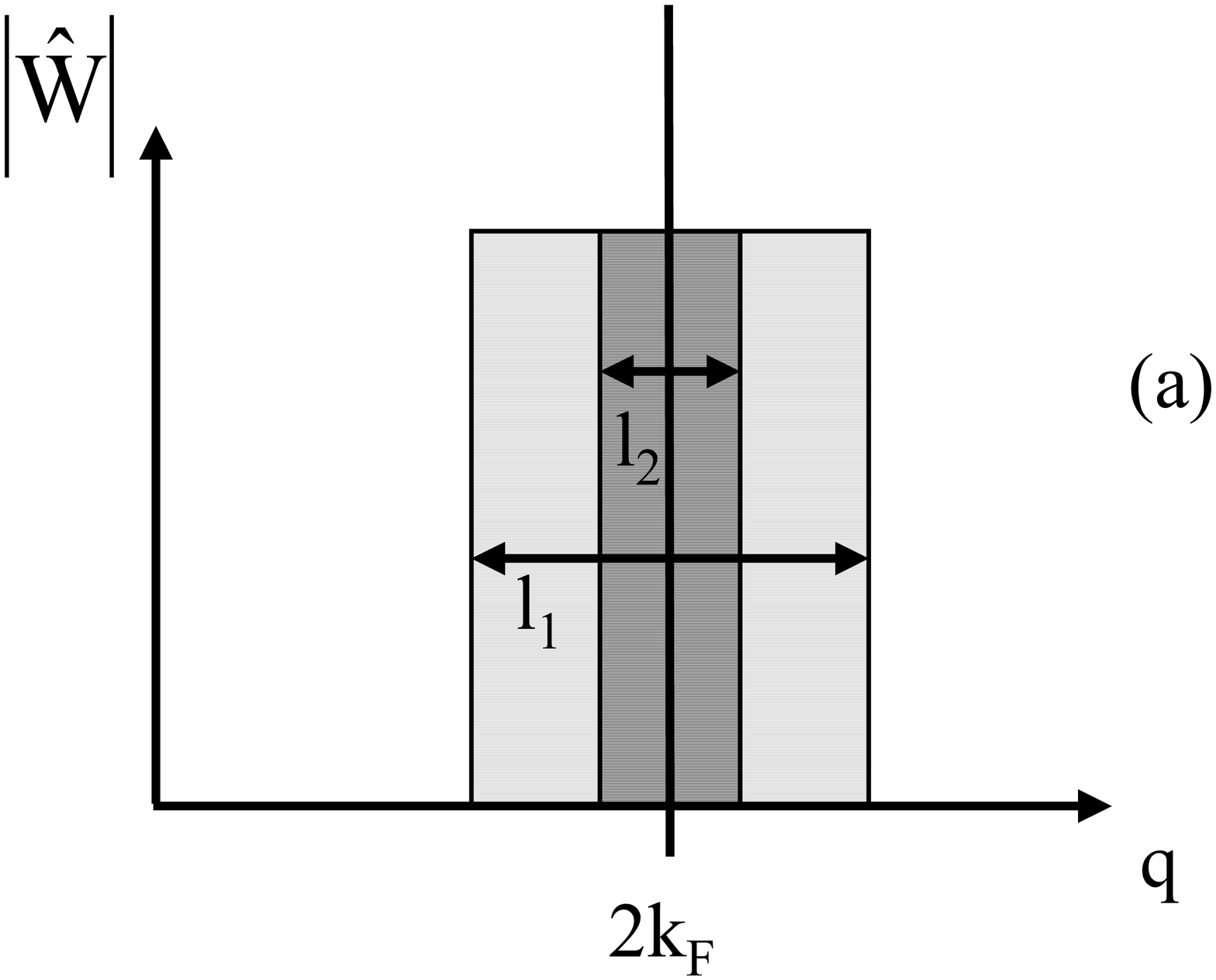}
                \includegraphics[width=\figwidth]{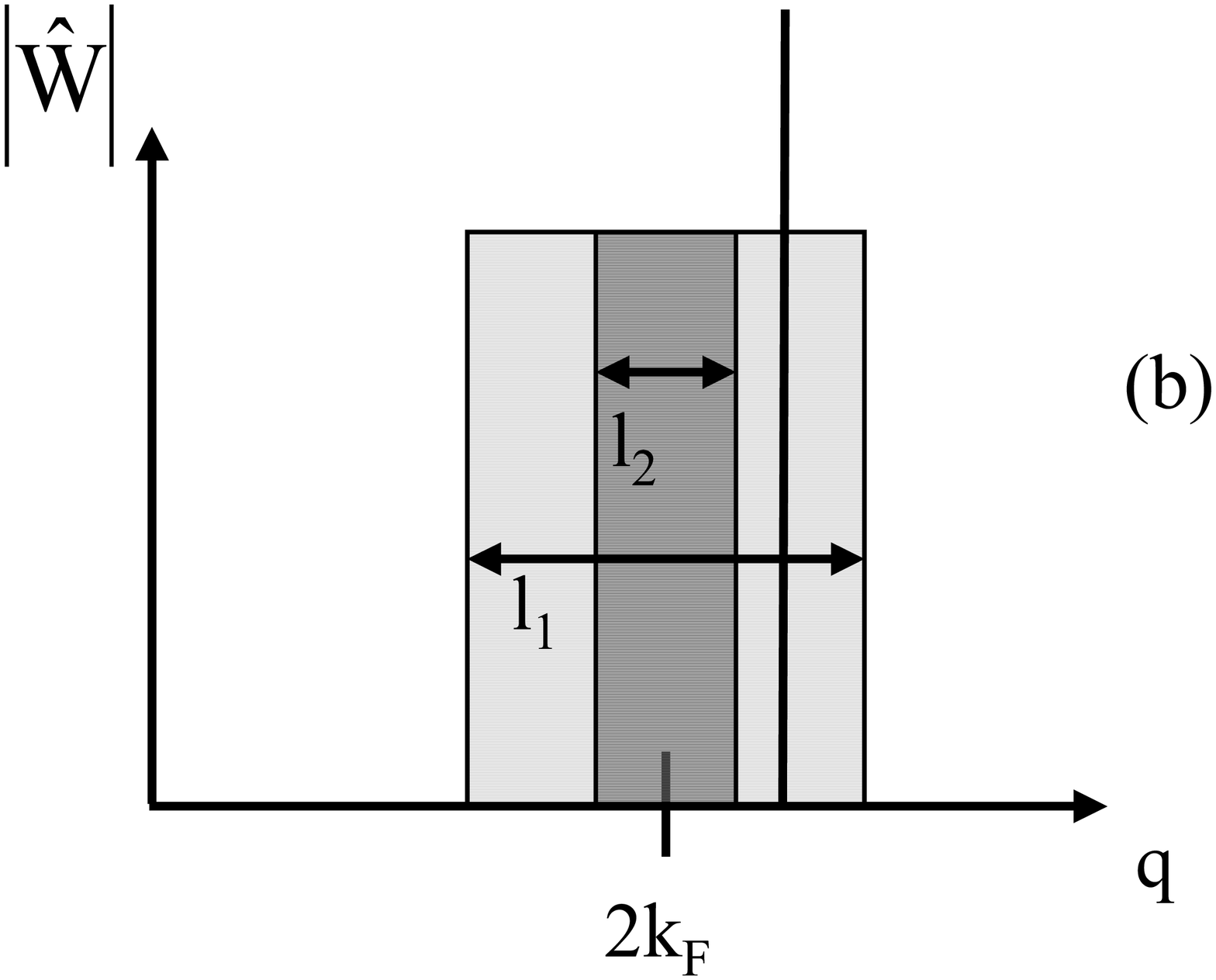}}
\centerline{\includegraphics[width=\figwidth]{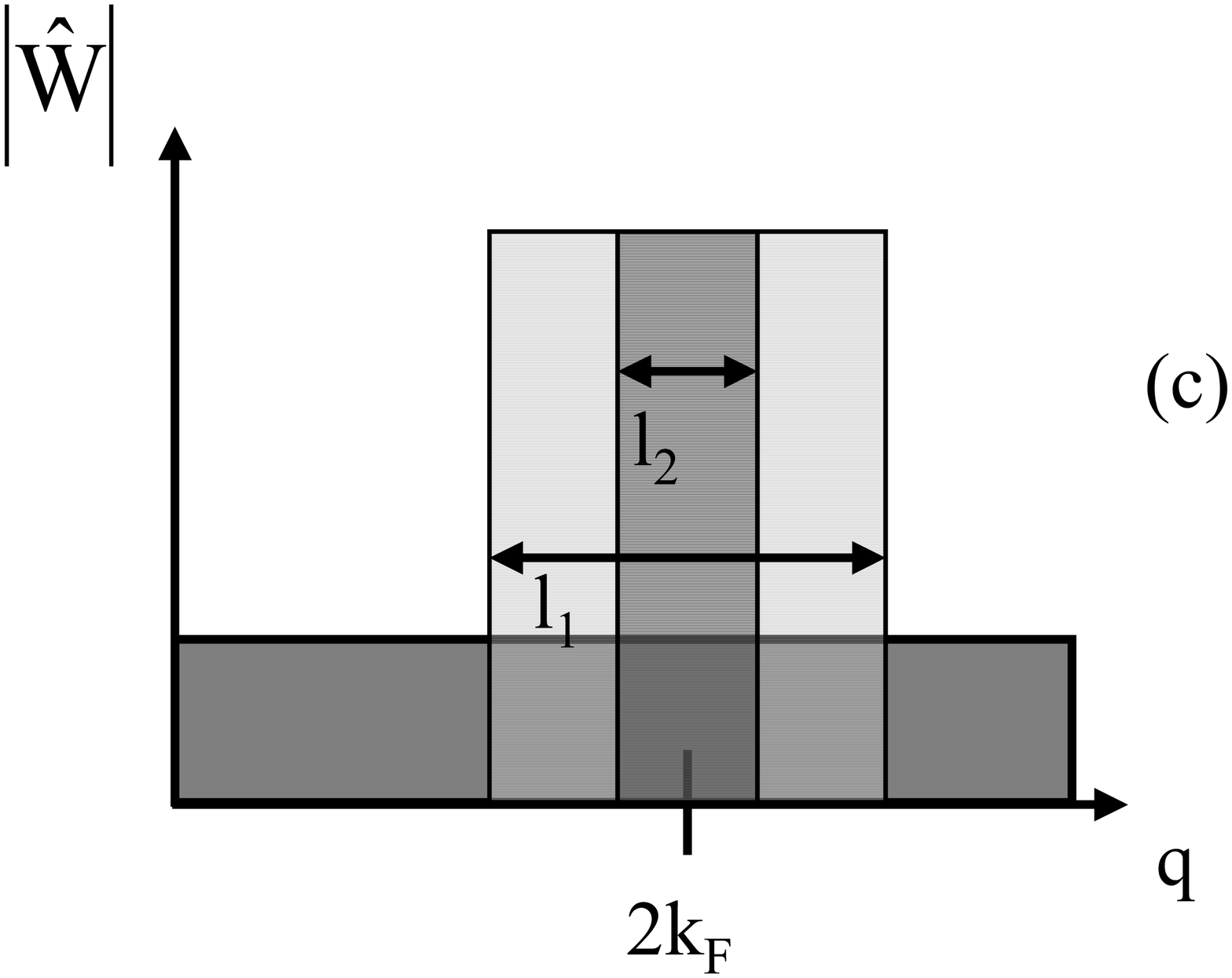}
               \includegraphics[width=\figwidth]{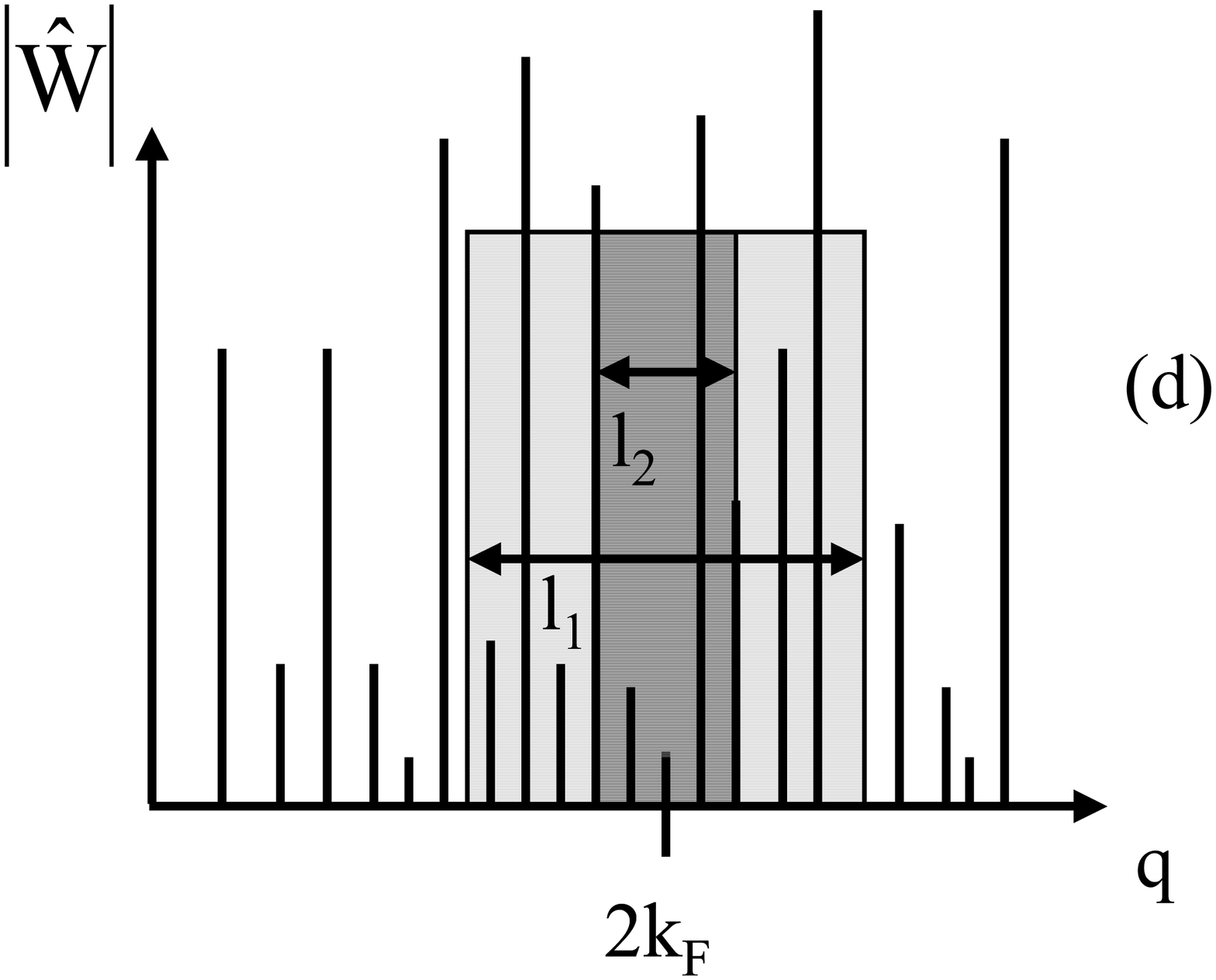}}
\caption{\label{fig:window} Schematic view of the renormalization
procedure. The function $J$ (see text) filters the Fourier components
$\hat W(q)$ of the potential and only keep those in a narrowing window around
the Fermi level ({\it i.~e.} for $q=2k_F$) as the scale increases. We
have displayed the windows for two different scales $l_1$ and
$l_2>l_1$. For the periodic
case the potential has only one peak. If the potential is
commensurate with $2 k_F$ (a), the
peak is inside the windows at all scales and the renormalization never stops.
If the potential is incommensurate (b) it only acts if the peak
remains in the windows
(which is the case for $l_1$ but not at the larger scale $l_2$).
For the disorder (c) case the potential has component at all $q$ thus
the strength of the potential efficient
in the renormalization is directly proportional to the size of the window.
For the quasiperiodic case (d) the potential has a complicated peaks
structure that gives rise to a novel behavior for the MIT.}
\end{figure*}

To study the influence of the potential $W$, we use a standard
RG approach by analyzing perturbatively the renormalization
of the correlation function (\ref{eq:correlationboso}) computed with
the full action of the system. First, note that via a redefinition of
the bosonic field $\phi$~:
%
%
\begin{equation}
\tilde\phi(x)=\phi(x)-{K \over u} \int^x dy\, W(y)
\mbox{,}
\end{equation}
%
%
the term proportional to $\partial_x\phi$ in
(\ref{eq:hwfinal}) can be absorbed in the quadratic part of the
action which thus writes~:
%
%
\begin{equation} \label{S0tilde}
S_0={1 \over 2 \pi} \int d{\bf r}\:
\left[{1\over uK}               (\partial_\tau \tilde\phi)^2 +
\left({ u \over K}\right) (\partial_x     \tilde\phi)^2\right]
\mbox{.}
\end{equation}
%
%
Introducing the Fourier components by~:
%
%
\begin{equation}
W(x)=\lambda\, \sum_Q {\hat W(Q)} e^{iQx} \mbox{ , }
\end{equation}
the potential part of the action reads~:
%
%
\begin{eqnarray}
S_W &=& {g \over u (2\pi \alpha)^2}  \sum_Q {\hat W(Q)}  \times \\
&&\int d{\bf r} \:
e^{i\left(2\phi(x,\tau) + Q^- x \right)} +
e^{i\left(-2\phi(x,\tau) + Q^+ x \right)}
\nonumber
\mbox{,}
\end{eqnarray}
where $Q^\pm = Q\pm 2k_F$ and $g=2\pi \alpha \lambda$.
Treating $S_W$ perturbatively and imposing that the asymptotic behavior
(\ref{decaying}) is unchanged when varying the cut-off $\alpha$
leads to the renormalization of the parameter $K$ and of the Fourier
components of the potential $W$.
The procedure is detailed in Appendix~\ref{ap:rgpot}. The RG
equations are given by~:
\begin{eqnarray}
{dK\over dl}&=&-K^2 \,\Xi(l) \mbox{,} \label{eqRG1}\\
{dy_Q\over dl}&=&(2-K) \, y_Q \label{eqRG2}
\mbox{,}
\end{eqnarray}
with~:
\begin{equation} \label{eq:fonxi}
\Xi(l)= {1 \over 2} \sum_Q y_Q^2
\left[ \mbox{J}(Q^+ \alpha(l)) + \mbox{J}(Q^-\alpha(l)) \right]
\mbox{,}
      \label{Gl}
\end{equation}
where the $y_Q=\lambda \alpha \hat{W}(Q)/u$ are the dimensionless
Fourier components of $W$ and $l$ is the scale factor defined by
$\alpha(l)=\alpha(0) e^l$ where $\alpha(0) = \alpha$ is proportional
to the original
lattice spacing $a$. In (\ref{Gl}), $\mbox{J}$  is a function  whose
precise form depends on the cut-off procedure used to eliminate the
short distance degrees
of freedom (see Appendix \ref{ap:rgpot}).
Different kind of functions
are considered below,
but one typically has~:
\begin{eqnarray}
\mbox{J}(Q^\pm \alpha(l))&\simeq& 1 \:\:\mbox{for} \:\:
\alpha(l)<1/Q^\pm  \label{eq:cut-offbis}\\
     &=& 0
\:\:\mbox{otherwise}
\label{eq:cut-off}
\mbox{.}
\end{eqnarray}
Physically, this means that the renormalization is equivalent to an
investigation of
the low-energy properties in a window around $2k_F$ in the reciprocal
space whose width
is proportional to $e^{-l}$. Thus, the {\it full} Fourier landscape
of the potential determines
the scaling of the parameters $K$ and $y_Q$. This is summarized in
Fig.~\ref{fig:window}.
%
%
%
%
\section{Critical properties}
\label{Critical_properties}

In this section, we discuss the relevance of the interactions for different
potentials. In other words, we analyze the possibility of phase transition
when varying the strength of the interactions.

\subsection{Periodic potentials} \label{sec:periodic}

This is the simplest case since the potential involves only one
harmonic~:
\begin{equation} \label{eq:periodsimp}
W(x)=\lambda \cos(Q\,x)
\mbox{.}
\end{equation}
Inserting the form (\ref{eq:periodsimp}) in (\ref{eqRG1}-\ref{eqRG2}) leads to
the following flow equations \cite{giamarchi_umklapp_1d}~:
\begin{eqnarray}
{dK\over dl}&=&-K^2 y^2 \left[\mbox{J}(Q^+\alpha(l)) +
\mbox{J}(Q^-\alpha(l)) \right]\mbox{,} \label{eq:flowakt}\\
{dy\over dl}&=&(2-K) y \label{eq:flowbkt}
\mbox{,}
\end{eqnarray}
where $y=\lambda \alpha /u$. Two different behaviors have to be
distinguished depending on
whether $\pm Q$ coincides with $2k_F$ or not.

\subsubsection{Incommensurate case ($Q^+\neq 0$ and  $Q^-\neq 0$)}

In this case,  there always exists a scale $l^*$ such that
$\mbox{J}(Q^\pm \alpha(l^*))=0$, as shown on Fig.~\ref{fig:window}.
At this scale the renomalization of $K$ essentially
stops. $K$ thus converges towards a fixed point $K^*$, and
the potential $W$ is irrelevant. The system remains a
Luttinger liquid with gapless
excitations\cite{giamarchi_umklapp_1d,giamarchi_mott_shortrev}
and the correlation functions decay with an effective exponent $K^*$.
This can be understood by the fact that, in this case, the Fermi
level does not lie in the
gap opened by the periodic potential $W$.

\subsubsection{Commensurate case ($Q^+=0$ or $Q^-=0$)}

Suppose for instance that $Q=2k_F$. In that case
$\mbox{J}(Q^-\alpha(l))=1$ at all scales and
the renormalization of $K$ cannot be stopped, as shown on
Fig.~\ref{fig:window}. The potential is commensurate
and would, for non interacting electrons ($K=1$) open a gap at the Fermi level.
For interacting particles, the effect of the potential is given by the
flow (\ref{eq:flowakt}-\ref{eq:flowbkt})
which is now a simple Beresinskii-Kosterlitz-Thouless flow
\cite{Berezinskii1,Berezinskii2,kosterlitz_modele_xy}.
Thus, for $K > 2$, the potential $W$ is irrelevant, and the system remains a
gapless Luttinger liquid. For $K<2$, the potential is relevant and the
system flows to
strong coupling regime. The strong coupling fixed point is
not reachable by the
perturbative flow but, since the system is
described in this case by a simple sine-Gordon action, we know, from
other methods, the physical properties of this phase
\cite{emery_revue_1d,schulz_houches_revue,voit_bosonization_revue}. A gap
opens in the spectrum between the ground state and the first excited state. An
estimate for this gap can be obtained from the RG analysis. If
$l^*$ denotes the lengthscale at which $y \sim 1$, the gap is given by~:
\begin{equation}
\Delta \propto e^{-l^*}
\mbox{.}
\end{equation}
In the case where $y \ll (2-K)$ one can neglect the
renormalization of
$K$ in (\ref{eq:flowbkt}) and we obtain~:
\begin{equation}
\Delta \sim y^{1\over 2-K}
\mbox{.}
\end{equation}
Note that for the non interacting point ($K=1$), one recovers the linear
scaling of $\Delta$ with respect to the strength of the potential
$\lambda$ expected by
the first order perturbation theory.
This MIT induced by the interaction is known as the Mott
transition.
Let us emphasize that this critical value $K_c=2$, separating a metallic phase
$(K>K_c)$ from an insulating one $(K<K_c)$, corresponds to attractive
interactions between
fermions.

The later discussion on the single harmonic case still holds for a
potential with a finite
number of Fourier components. The possibility of an insulating regime
is then offered when the Fermi level is in one of the gaps  opened by
the various
frequencies of the
potential. The most interesting situation thus arises when the Fourier spectrum
of the potential
is dense or continuous. Such potentials can be encountered in several
physical systems
but we focus here on two of them~: the random  and the quasiperiodic
potentials.

\subsection{Disordered potentials}

Let us consider a potential provided by a random variable with
uniform probability whose Fourier components satisfy~:
\begin{equation}
\overline{\hat W^*(Q)\hat W(Q')}=\lambda \,\delta_{QQ'}
\mbox{.}
\end{equation}

\noindent Using the general expressions  (\ref{eqRG1}) and
(\ref{eqRG2}) one obtains~:
\begin{eqnarray}
{dK\over dl}&=&-{ K^2 \, y^2\over 2} \sum_Q
\left[ \mbox{J}(Q^+ \alpha(l)) + \mbox{J}(Q^-\alpha(l)) \right]
\mbox{,} \label{K_dis}\\
{dy\over dl}&=&(2-K) \, y
\label{y_dis}
\mbox{.}
\end{eqnarray}
In Eq. (\ref{K_dis}), the sum actually stands for an integral so one has~:
\begin{widetext}
\begin{equation}
\sum_Q \left[ \mbox{J}(Q^+ \alpha(l)) + \mbox{J}(Q^-\alpha(l)) \right]
\rightarrow
\int dQ \left[ \mbox{J}(Q^+ \alpha(l)) + \mbox{J}(Q^-\alpha(l)) \right]
\sim {1 \over \alpha(l)} \sim e^{-l}
\mbox{.}
\end{equation}
\end{widetext}
This means that the renormalization of $K$  is directly proportionnal to the
window width around $2k_F$ at scale $l$.
In the limit of weak disorder the RG equations
$(\ref{K_dis}-\ref{y_dis})$ can be integrated
neglecting the renormalization of $K$~: $y(l)=y(0) \ e^{(2-K)\,l}$.
One then has~:
\begin{equation}
{dK\over dl}=- C K^2  e^{(3-2K)\,l}
\mbox{,}
\end{equation}
where $C$ is a constant. One deduces the existence of a critical point at
$K_c=3/2$. Our approach thus genera\-lises the RG treatment specific to the
disorder case \cite{apel_spinless,giamarchi_loc}.
For $K<K_c$, the
system is insulating {\em for any filling} whereas it can
be metallic for sufficiently attractive interaction. Note that for
$K=1$, the system is an
insulator as expected for a one-dimensional disordered non interacting system.
By contrast to the  periodic case, this transition point is independent
of $k_F$. Physically,
this can be understood invoking the proximity of several localized
states (for $K=1$)
at arbitrarily short distance that can couple to each other if the
interaction is attractive
enough.
It is interesting to remark that the critical value in the disordered
case ($K_c=3/2$) is smaller than
in the periodic case ($K_c=2$). This means that the localization
induced by the disorder
is more easily destroyed by attractive interactions than the one
induced by a finite width gap.
In this context, it is clear that  dense Fourier spectrum are
able to provide
rich outstanding situations where the position of the Fermi level
could, in principle, determine the value  of a critical point.

\subsection{Quasiperiodic Potentials}

\begin{figure}
\centerline{\includegraphics[width=\largewidth]{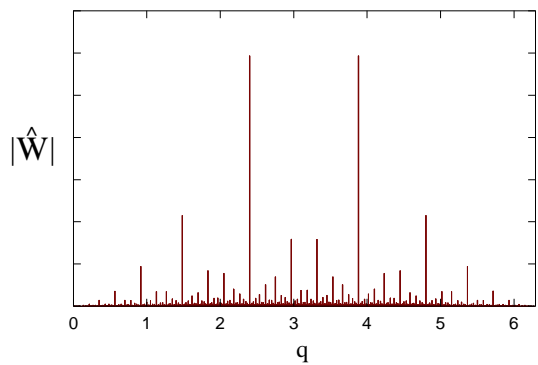}}
\caption{\label{TFGM}
Fourier transform of the $15^{th}$ approximant of the Golden Mean
(Fibonacci) potential (610 sites per unit cell).}
\end{figure}

\begin{figure}
\centerline{\includegraphics[width=\largewidth]{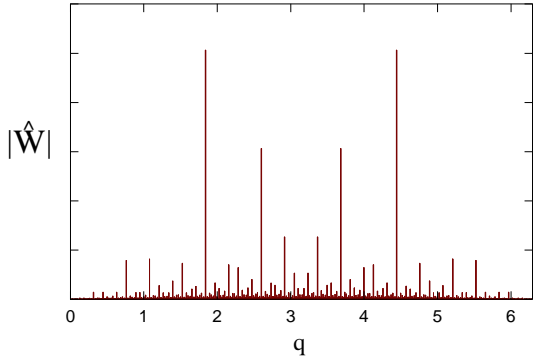}}
\caption{\label{TFSM}
Fourier transform of the $10^{th}$ approximant of the Silver Mean
potential (1393 sites per
unit cell).}
\end{figure}

To analyze such situations, we consider quasiperiodic potentials.
Several types of potentials can play this role but, as already explained,
the most interesting ones  are those that have a non trivial dense
Fourier spectrum.
Here, we focus on quasiperiodic potentials obtained by substitution
rules which lead to quasiperiodic effects even at the perturbative level.
Among all possible choices, we consider the simplest one, known as precious
mean potentials, that are given by the following iterative scheme~:
\begin{equation} A \rightarrow A^{k}B,\hspace{2.ex} B\rightarrow A
\mbox{.}
\end{equation}
We associate, to each site, a diagonal potential that can take
two discrete values
$W_A=+\lambda/2$ or $W_B=-\lambda/2$.
Let us note that a global shift of the $W_i$ always allows to deal with a
zero-averaged potential so that we can set $\hat W(0)=0$.
For $k=1$, one recovers the famous Fibonacci chain associated to the Golden
Mean, for $k=2$ one has the Silver Mean sequence, etc. We give in the
Appendix~\ref{ap:precious}
a brief description of these sequences and a detailed calculation of
their Fourier tranform.
As it can be inferred  from Fig.~\ref{TFGM} ($k=1$) and 
Fig.~\ref{TFSM} ($k=2$),
the Fourier spectrum is dense in $[0, 2\pi]$ in the quasiperiodic limit,
and has a multifractal structure \cite{Peyriere}.
As in the periodic case, we have to consider several situations since
we expect a strong
dependence of the physical properties with respect to the position of
the Fermi level.
It is clear that if there exists a wave vector $Q\simeq 2k_F$ such that  $\hat
W(Q)$ is large compared to
the other Fourier components, the electrons behave as if they were
embedded in a periodic potential, whereas in the opposite case, {\it i.~e.}
the Fermi level lies in
a very small gap, one could expect another behavior. Of course, the
notion of far or close
from a gap has only a sense once we have specified the maximum scale
$l_{max}$ up to which
we study the RG equations.  In our case, since we only consider approximant
(arbitrarily large), $\alpha(l_{max})$ is given by the inverse of the typical
distance between gaps.
Beyond this scale, the RG flow is no more sensitive to the
precise structure of the Fourier spectrum. One can then encounter
three situations~:

\subsubsection{$\hat W(2k_F)$ is large}

\begin{figure}[h]
\centerline{\includegraphics[width=\largewidth]{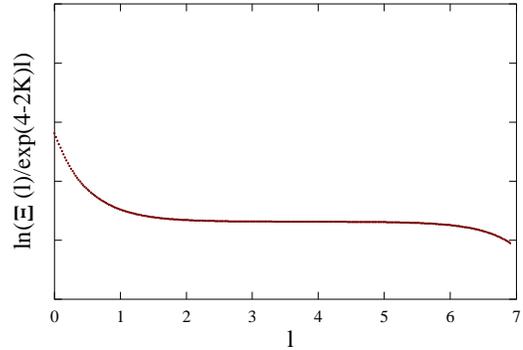}}
\caption{\label{grandpic}
Behavior of $\Xi$ when the Fermi level $2k_F=2.4$ is close to a
dominant peak (Fibonacci
potential).}
\end{figure}

There is an harmonic of $W$  at  $Q\simeq 2k_F$ such that $\hat W(2
k_F)$ is large
compared to the other Fourier components. In other words, $\hat W(Q)$ opens a
finite gap that
dominates the low-energy
physics up to a scale given by $(Q-2k_F)^{-1}$. The function $\Xi$
that governs the flow of
$K$ is then completely dominated by this component and behaves as
$e^{(4-2K)\,l}$
(see  Fig.~\ref{grandpic}).
This actually defines the proximity of a ``dominant" peak.

In this case, one recovers a critical value $K_c=2$ corresponding to
that obtained in the periodic case, separating a metallic phase
($K>2$) from a
insulating one ($K<2$).

\subsubsection{$\hat W(2k_F)$ is small}
\label{small}

\begin{figure}[h]
\centerline{\includegraphics[width=\largewidth]{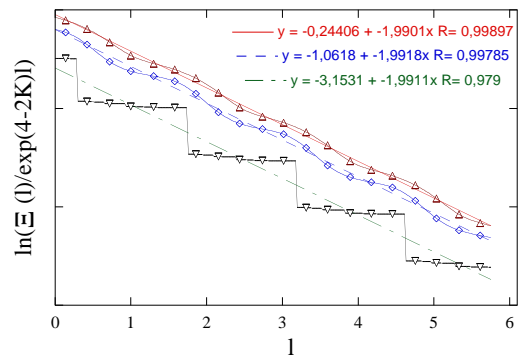}}
\caption{\label{Regulateur}
Behavior of $\Xi$ at half-filling ($2k_F=\pi$) for the Fibonacci
potential. We have displayed the results for three different
regulating function $\mbox{J}$~:
$\mbox{J}(x)=e^{-x^2}$$(\bigtriangleup)$,
$\mbox{J}(x)=1/(1+x^6)$ $(\diamond)$ and the step function
$\mbox{J}(x)=1$ if $x<1$, 0 otherwise $(\bigtriangledown)$.}
\end{figure}

The Fermi level lies far from the main gaps. An example of such
a situation is obtained
at half filling $(2k_F=\pi)$. In this case, as shown in Fig.~\ref{Regulateur},
$\Xi(l)\sim e^{(4-2K-2)\,l}$ for any kind of regulator.

\begin{figure}[h]
\centerline{\includegraphics[width=\largewidth]{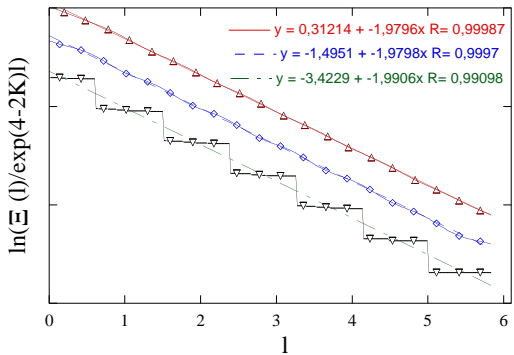}}
\caption{\label{Regulateur2}
Behavior of $\Xi$ at half-filling ($2k_F=\pi$) for the silver mean
potential. We have displayed the result for three different
regulating function $\mbox{J}$~:
$\mbox{J}(x)=e^{-x^2}$$(\bigtriangleup)$,
$\mbox{J}(x)=1/(1+x^6)$ $(\diamond)$ and the step function
$\mbox{J}(x)=1$ if $x<1$, 0 otherwise $(\bigtriangledown)$.}
\end{figure}

Of course, this is not strictly speaking an exponential decrease and 
there is, in
particular, some oscillations (in log-log) that are reminiscent of 
the multifractality
of the Fourier spectrum. However, one can fairly approximates $\Xi(l)$ by
$e^{2(1-K)\,l}$ and we obtain  a critical point at $K_c=1$.
This critical value
corresponds to the non interacting point $V=0$.
This means that when
the Fermi level lies in zero width gap (small peaks) the slightest
attractive interaction
allows the system to become metallic whereas it is insulating at $V=0$.
   From a spectral point of view, this means that an attractive interaction
close the gap and allows for arbitrarily low-energy excitations. This
point will be discussed more in Sec.~\ref{sec:beyond}.
We would like to stress that other positions of the Fermi level ($2k_F=0.5$ for
Fibonacci for example) also gives $K_c=1$, and that similar
situations occurs for other
precious mean potentials (see Fig.\ref{Regulateur2}).

%
%
%
%
%
%
\subsubsection {$\hat W(2k_F)$ is intermediate}
\begin{figure}[h]
\centerline{\includegraphics[width=\largewidth]{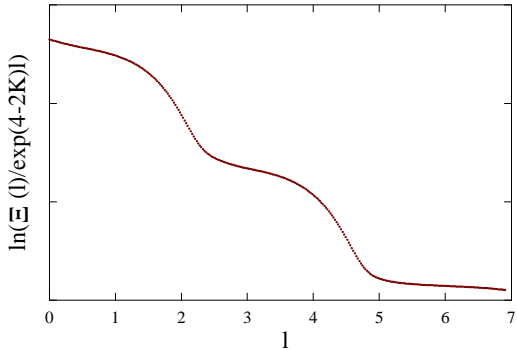}}
\caption{\label{moyenpic}
Behavior of $\Xi$ when the Fermi level $2k_F=0.5$ (Fibonacci potential)
is in an intermediate
situation.}
\end{figure}

Finally, there are intermediate situations for which $\Xi(l)$
does not have an exponential behavior, even approximately.
In these cases, it is impossible to simply extract a critical exponent and
one needs a non perturbative treatment of the problem to determine a
possible transition point
(see Fig.~\ref{moyenpic}). Note that in this case, one still
observes the oscillations.

\section{Physical consequences} \label{Transport}

In addition to the phase diagram and the critical properties,
determined in the previous section, the RG allows to extract many
physical properties. In the metallic regime $K > K_c$ (where $K_c=2$ for
periodic potential, $K_c=3/2$ for disordered potentials and $K_c=1$ for special
filling factor in quasiperiodic potential), the system flows to weak
coupling so the RG can be
used at arbitrary scales. For the insulating side $K < K_c$ the
potential flows to strong
coupling. One can thus use the RG to obtain the properties up to the
lengthscale
$l^*$ such that the potential at this lengthscale is of order one. The behavior
beyond this lengthscale (or below the correponding energy scale) can
not be accessible by
the RG and should be treated by other (non perturbative) methods.
This point will be discussed
in Sec.~\ref{sec:beyond}.

\subsection{Transport properties}

Both the d.~c. and a.~c. transport properties of the system can in
principle be extracted from the Kubo formula. The fermionic current
operator can easily be written using the bosonic variables and reads~:
\begin{eqnarray}
j &=& v_F(\psi_R^\dagger\psi_R - \psi_L^\dagger\psi_L) \\
       &=&  u K \Pi = \partial_\tau\phi \label{eq:current}
\mbox{.}
\end{eqnarray}
Thus the conductivity is simply given by the correlation function~:
\begin{equation} \label{eq:conduc}
\sigma(\omega) = \frac{i}{\omega}\left[
          \frac{2 u K}\pi + \chi(\omega) \right]
\mbox{.}
\end{equation}
$\chi(\omega)$ is the retarded current-current correlation function~:
\begin{equation} \label{eq:retard}
\chi(\omega) = \langle j; j \rangle_\omega =
       - \frac{i}L\int dx \int_0^\infty dt \langle [j(x,t),j(0,0)] \rangle
e^{i\omega t}
\mbox{,}
\end{equation}
where $j$ is given by (\ref{eq:current}), and $L$ is the size of the system.
In the absence of $H_W$, since the Hamiltonian
is quadratic in the bosonic variables (see
(\ref{eq:ham_boso_texte})), (\ref{eq:retard})
is trivially computed and the conductivity is given by~:
\begin{equation} \label{freecon}
\sigma(\omega) = u K \left[\delta(\omega) + \frac{i}\pi {\cal P}
\frac1\omega \right ]
\mbox{.}
\end{equation}
The system is a perfect conductor and $uK$ plays the role of the
standard plasma frequency.
Computing fully (\ref{eq:conduc}) in the presence of $H_W$ is of
course impossible, but one can use an hydrodynamic approximation
\cite{giamarchi_umklapp_1d,giamarchi_attract_1d,giamarchi_mott_shortrev},
using the so-called memory function formalism
\cite{gotze_fonction_memoire} which is well adapted to the
one-dimensional situation.
For completeness, we sketch the main steps of the method
in Appendix~\ref{ap:memoire}. The conductivity thus writes~:
\begin{equation} \label{eq:condmemtext}
\sigma(\omega) = \frac{i2uK}\pi \frac1{\omega + M(\omega)}
\mbox{,}
\end{equation}
where the function $M$ is perturbatively given by (see
Appendix~\ref{ap:memoire})~:
\begin{equation} \label{eq:appro}
M(\omega) = \frac{[\langle F;F\rangle_\omega^0 - \langle F;F
\rangle^0_{\omega=0}]/\omega}{-\chi(0)}
\mbox{.}
\end{equation}
The $F= [j,H]$ operator takes into account the fact that the current is not a
conserved quantity and $\langle F;F \rangle_\omega^0$ stands
for the retarded correlation function of the operator $F$ at
frequency $\omega$ computed in the {\it absence} of the scattering potential
($H_W = 0$).
Using (\ref{eq:hwfinal}) leads to~:
\begin{equation} \label{eq:f}
F(x) =  \frac{2u}{\pi \alpha} W(x)   \sin(2\phi(x)-2k_F x)
\mbox{.}
\end{equation}
The memory function can be easily computed for periodic potential
with a single
harmonic\cite{giamarchi_umklapp_1d,giamarchi_mott_shortrev} or for an
uncorrelated
disordered potential. In order to get
the behavior beyond the simple perturbation, it is necessary to
couple (\ref{eq:appro}) with the RG calculation
\cite{giamarchi_umklapp_1d,giamarchi_mott_shortrev}. The
RG is iterated until the frequency or temperature is comparable to the
renormalized cut-off. The memory function (\ref{eq:appro})
with the {\it renormalized} parameter gives then the conductivity.
Using the expression
of the current (\ref{eq:f}) one can compute the scaling dimension of $M$.
  From (\ref{eq:appro}) one gets~:
\begin{eqnarray}
M(\omega) &=& \frac1\omega \int dx \int_0^\beta d\tau
\left(e^{i\omega \tau} -1 \right)  \times \nonumber\\
& & \sum_Q |\hat W(Q)|^2 [ e^{i Q_+ x} +
e^{i Q_- x} ] \left(\frac{\alpha}{r}\right)^{2K}
\label{eq:scalingm}
\mbox{.}
\end{eqnarray}
Eq. (\ref{eq:scalingm})
is essentially the expression of $\Xi$ appearing in the RG calculation
(\ref{eq:fonxi}). The
differences are
just trivial scaling factors.
Thus if the function $\Xi$ varies as:
\begin{eqnarray}
\Xi(l) &\propto& e^{(4 - 2K - \mu)l}
\mbox{,}
\end{eqnarray}
%
%
where $\mu$ is a real number, then from (\ref{eq:scalingm}) one obtains~:
\begin{eqnarray} \label{eq:memscaling}
M(l) &\propto& e^{(3 - 2K - \mu)l}
\mbox{.}
\end{eqnarray}
The function $\Xi$ thus gives directly the conductivity.
Stopping the RG flow when $\alpha(l^*) \sim \max(\omega,T)$, and
replacing $M(l^*)$ in (\ref{eq:condmemtext}) gives both the d.~c. and
a.~c. conductivity.

\begin{figure}[h]
\centerline{\includegraphics[width=60mm]{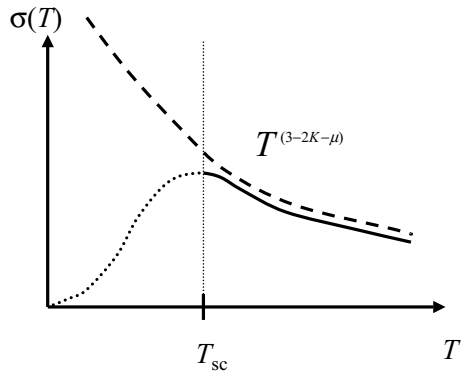}}
\caption{\label{fig:conduc} Schematic dependence of the temperature
dependence of the conductivity. Above the MIT  point $K > K_c$
the conductivity (dashed line) is power-law like
with an exponent depending on the interactions. On the insulating
side this behavior remains valid in the weak coupling regime, {\it
i.~e.} for temperatures
above the crossover scale $T_{c}$ (see text).}
\end{figure}

Simplified expressions can be obtained for very weak
potential, and high
enough temperatures or frequencies. In that case, one can neglect the
renormalization
of $K$ in the RG flow.
Since $M$ is small in this regime, the conductivity is given by~:
\begin{eqnarray}
\sigma(\omega,T=0) &\sim& \frac{i}{\omega} -
\frac{M(\omega,T=0)}{\omega^2} \mbox{,}\\
\sigma(\omega=0,T) &\sim& \frac1{M(\omega=0,T)}
\mbox{,}
\end{eqnarray}
leading to~:
%
%
\begin{eqnarray}
\sigma(\omega) &\propto& \omega^{2K-5} \qquad \sigma(T) \propto
T^{3-2K} \quad \text{(commensurate)}, \nonumber\\
\sigma(\omega) &\propto& \omega^{2K-4} \qquad \sigma(T) \propto
T^{2-2K} \quad \text{(disordered)},   \nonumber\\
\sigma(\omega) &\propto& \omega^{2K-3} \qquad \sigma(T) \propto
T^{1-2K} \quad \text{(quasi. for $K_c=1$)} \nonumber
\end{eqnarray}
%
%
A sketch of the conductivity is shown in Fig.~\ref{fig:conduc}.

In the metallic regime $K> K_c$ there is, in addition, a Drude peak,
whose weight
is given by the renormalized Luttinger liquid parameters as $D = u^* K^*$.
In the insulating regime, since the RG can only be pushed until the
scale $l_{c}$ for which
$\Xi(l_{c}) \sim 1$ these expressions, and their generalization when
the renormalization of
$K$ is taken into account
\cite{giamarchi_umklapp_1d,giamarchi_mott_shortrev}, are only valid for
temperatures and frequencies smaller than $\alpha e^{l_{c}}$.
This corresponds to the Mott gap for the commensurate system and the
pinning frequency (inverse localization
length) for the disordered one as will be discussed in more details
in Sec.~\ref{sec:beyond}.

For a mesoscopic system of size $L$, it is often more interesting to
compute the conductance
$G$ of the system as a function of the size $L$. This is in general a
much more complicated
calculation, specially for interacting systems for which is is difficult to use
the Landauer formula. However, it is possible to extract
the scaling behavior from the memory function as well
\cite{ogata_wires_2kf,giamarchi_moriond}. Indeed, using $I = G V$
($V$ being the electrostatic potential) the conductance is given by~:
%
%
\begin{widetext}
\begin{equation} \label{eq:con}
G = \lim_{\omega\to 0} \frac1{i(\omega+i\delta)}
\left[\left(-\frac1{L}\int_{-L/2}^{L/2} dx'
\int dt' e^{i(\omega+i\delta)(t-t')}
\langle j(x=0,t);j(x',t') \rangle\right) - \frac{D}{L}\right]
\mbox{,}
\end{equation}
\end{widetext}
%
%
\mbox{ }\\
which, rewritten in term of the conductivity, gives~:
\begin{eqnarray}
G &=& \int dq \frac1L \int_{-L/2}^{L/2} dx e^{i q x} \sigma(\omega\to 0,q) \\
        &=& \int dq \frac{\sin(qL/2)}{qL/2} \sigma(\omega\to 0,q)
\mbox{.}
\end{eqnarray}

The cardinal sine can be considered as a simple cut-off, restricting
the integral over $q$.
The conductance (in units of $e^2/h$) is thus simply given by~:
\begin{eqnarray} \label{eq:condapp}
G \simeq \int_{1/L}^{1/L} dq \sigma(\omega\to 0,q)
\mbox{.}
\end{eqnarray}
%
%

%
%
%
%

When $H_W=0$ the conductance is simply $G=K$. Note that the fact
that conductance is not $G=1$ for the pure system but depends on the
interactions is in general an artefact, that has to be corrected
depending on the
system\cite{safi_pure_wire}. However, since
we focus here on the effect of a scattering potential, this is not
important for our purpose.

A simple way to compute the conductance in (\ref{eq:condapp}) is
again to iterate the RG flow
until the cut-off is of the order of the size of the system. Up to
that point the $q$ dependence
can be neglected and $\sigma(\omega) \sim 1/M(\omega)$.  The
corrections to the conductance are thus given by~:
\begin{eqnarray}
\Delta G \propto \frac1L \frac1{M(l = \ln(L/\alpha))}
\mbox{.}
\end{eqnarray}
Thus, with the scaling (\ref{eq:memscaling}), one has~:
\begin{eqnarray} \label{eq:scacond}
\Delta G \propto L^{2K + \mu - 4}
\mbox{.}
\end{eqnarray}

For the disordered case, in the absence of any renormalization of the
disorder, (\ref{eq:scacond}) leads to a variation of the resistance
$\Delta R=1/\Delta G \propto L$ in the absence of
interactions which is nothing but
Ohm's law. Interactions and
renormalization of the parameter $K$ by disorder change this scaling
\cite{giamarchi_loc,giamarchi_moriond}.
Of course the renormalization of disorder also affects the exponents
through the renormalization equations (\ref{eqRG1}-\ref{eqRG2}). The
faster increase of
the resistance with size is the
sign of Anderson localization.
As discussed before, the RG results can only be used if the
renormalized scattering potential
remains small compared to 1. This means that $\Delta G < e^2/h$. To
go beyond, one
needs to know the strong coupling fixed point. The lengthscale at
which this happens
is of course the Mott length for the commensurate potential and the
localization length
for the disorder as we now discuss in more details.
%
%
\subsection{Beyond the RG} \label{sec:beyond}
%
%
In the metallic phase the RG gives the full physics of the system, up to
arbitrarily low energy scales.
On the insulating side, on the other hand, the RG flows to strong coupling.
It thus defines a scale $l_{c}$ for which the function
$\Xi$ becomes of order one. Below the lengthscale $L \sim \alpha e^{l_c}$ the
RG still gives directly the physical properties, as shown for example
on Fig~\ref{fig:conduc}.
Note that this lengthscale can be quite large if the system is close
to the MIT or
for instance when the Fermi level is far from one of the main peaks.
Beyond this lengthscale a knowledge of the strong coupling fixed
point is in principle
necessary to describe the physics of the system. Fortunately many of
the properties
can still be inferred directly from the Hamiltonian. Let us examine
the various cases.

For the commensurate case, we know that the potential opens a gap in the
spectrum. Thus, we can relate the crossover scale to the gap by~:
\begin{equation}
\Delta/W \sim e^{-l_{c}} \sim \frac\alpha\xi
\mbox{.}
\end{equation}
Thus the RG gives directly the gap. The crossover scale is in this case the
so-called Mott length $\xi$ above which the correlation functions have
exponential decay. Similarly the conductivity decreases
exponentially for temperatures below the gap.

In the disordered case, the situation is more subtle. The disorder
does not open a gap, but we know that if it is strong enough,
wavefunctions are exponentially
localized, with a localization length of the order of the lattice spacing.
One can thus again relate the crossover length to the localization
length of the system by \cite{giamarchi_loc}~:
\begin{equation}
\xi \propto \alpha e^{l_{c}}
\mbox{.}
\end{equation}
Because of the exponential localization of the wavefunctions, $\xi$
also defines a scale below which most of the interaction effects stop
to be important.
\newpage
  Indeed the frequency dependence of the conductivity
become for $\omega < \omega_p = u/\xi$, $\sigma(\omega) \sim
\omega^2$ (up to log corrections) as for a non interacting system.

In the quasiperiodic case, the strong coupling fixed point is elusive
so we can make only educated guesses. In the noninteracting case ($K=1$)
for a point of the spectrum, the correlation functions decay as
a power-law. It is thus unlikely that for the interacting case the
strong coupling regime has
a characteristic lengthscale in a similar way than the commensurate
or disordered
system. Thus the most likely possibility is that
for the quasiperiodic case the length $\xi = \alpha e^{l_{c}}$
separates, two power-law regimes with different
exponents. Since for the noninteracting quasiperiodic
case, contrarily to the disordered case, the correlation functions are
still power-law, interactions are likely to still play a role even in
the strong coupling
regime. One can thus naively expect in that case that the exponent in
transport and
other correlation functions still depend on the interaction
strength, albeit probably in a different
way than in the weak coupling regime. This crossover is schematically shown
on Fig.~\ref{fig:conduc}.

\begin{figure}
\centerline{\includegraphics[width=\figwidth]{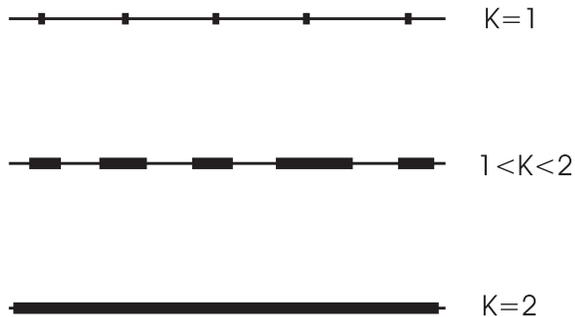}}
\caption{\label{fig:spectrum}
A possible evolution of the density
of states as a function of the Luttinger liquid parameter $K$. For
$K=1$ the spectrum is a set
of measure zero and gaps are present almost everywhere. For $1<K<2$
the small gaps
close as $K$ increases. For $K=2$ even the largest gaps are closed
since even a periodic potential is irrelevant above $K=2$.}
\end{figure}

The spectrum can also be inferred. For the
non interacting case $K=1$ it consists in an infinite set of zero width bands.
Clearly, the largest gaps must closed for $K=2$, as it is the case
for the periodic system.
Since there is, as shown in Section \ref{small}, a MIT at $K=1$ for
some filling
fractions, the smallest gaps should close for $K=1+\epsilon$ ($\epsilon\ll 1$).
We thus naturally expect larger and larger gaps to gradually close as
the interactions become more
and more attractive and the Luttinger parameter moves from $K=1$ to $K=2$.
Such an evolution of the spectrum as a function of $K$ is depicted in
Fig.~\ref{fig:spectrum}. It would be interesting to check this
scenario by numerical investigations.
%
%

%
%
\section{Conclusion}
\label{sec:conclusions}
%
%
We have studied in this paper a one-dimensional system of spinless
fermions submitted to
a quasiperiodic potential. Using a bosonization technique to treat
the interactions exactly and the RG approach we had introduced in
Ref.~\onlinecite{vidal_quasi_interactions_short},
we have investigated the effects of various types of quasiperiodic
potentials known as
precious potentials. We show that quasiperiodicity leads to a
novel class of MIT as a function of the strength of the interactions, since
for special filling factor the transition is pushed to the non
interacting point (insulator for repulsive interactions and
metallic for attractive ones).
We have determined the critical exponents and the associated
lengthscales and showed the
universality of the results for all types of precious potentials.
Our results are in good agreement with recent numerical investigations
\cite{Hida_quasi_spinless_DMRG,Hida_precious,eilmes_mit_twoparticles}.

We have also analyzed the transport properties such as the conductivity and the
conductance. These quantities behave as power-laws with respect to the
temperature (or with respect to the size for the conductance at $T=0$) with an
interaction-dependent exponent.
In the metallic regime, the RG flow converges towards a fixed point
which allows to extract
the full properties of the system. In the insulating regime, the
system flows to a strong
coupling regime at a lengthscale that we have determined.  Above this
lengthscale, it is
necessary to analyze the physics with non pertubative method.
Fortunately, one can still
estimate qualitatively the behavior of some of the quantities. For
instance, concerning the
spectrum one expects, as the interaction becomes attractive (or
equivalently as the Luttinger
parameter reached $K>1$), that the smallest gaps start closing until
$K=2$ where all gaps are
closed. Similarly for the temperature dependence of the conductivity
one expects
below the crossover scale $T_{c}$, a power-law dependence of
the conductivity. It would be interesting to
check the above proposals in the numerical solutions, both for spinless
\cite{Hida_quasi_spinless_DMRG} and spinfull
\cite{Hida_quasi_spinfull} systems. More
generally, the extension of these investigations to other types of
potentials such as the
Prouet-Thue-Morse or the paper-folding potentials
\cite{Davis_paperfolding} would be very useful since they also display a
complex (dense) Fourier
spectrum. Being able to tackle strong modulations would also allow
for a comparison with
potentials such as the
Harper potentials\cite{eilmes_mit_twoparticles} which at the
perturbative level are similar to simple periodic ones.

Several other questions are prompted by our study. The first one concerns
the temperature dependence of the conductivity.
The formula we derived assume that there are phase breaking
processes, so that the temperature acts indeed as a
cut-off in the RG \cite{giamarchi_umklapp_1d}.
The validity of this assumption has been recently
explicitly proven for the Mott (periodic) case\cite{rosch_conservation_1d}.
In that case, the phase breaking is provided by higher order
periodicity (higher order
umklapps). In the absence of such terms, the conductivity would
remain infinite.
It would be interesting to carry on the same type of memory matrix
approximation
for the quasiperiodic case.
Second, it would be interesting to know if a correlated disorder
could induce the same type of MIT as the one encountered here for the
quasiperiodic system. Indeed, this type of disorder is susceptible to
also produce a non
trivial Fourier spectrum and thus to have a critical $K$ that depends
explicitely on $k_F$.

To conclude, we address the question of experimental realizations to
directly observe
these effects. One could think about quasiperiodic chains in various devices.
Since in one dimension there is a direct equivalence between spinless
fermions and bosons
\cite{haldane_bosons,giamarchi_loc} it is possible to investigate the
physical properties of
quasiperiodic chain in Josephson junction arrays
\cite{fazio_josephson_junction_review}.
The advantage of such systems, besides the excellent control that one
can have on the
potential, is that we can  reach the attractive regime.
One could also realize a quasiperiodic chain using quantum dot arrays
\cite{kouwenhoven_quantum_dots_1d} or patterning of a quantum wire
\cite{tarucha_quant_cond,yacobi_quantumwire_conductance,depicciotto_qu 
antumwire_conductance}.
Finally, it would be interesting to check whether quasiperiodicity is
relevant to describe systems such as DNA for which there has been
recent transport
measurements\cite{tran_dna_optics,fink_dna_conductivity,kazumov_dna_supra}.

\begin{acknowledgments}
We would like to thank Cl. Aslangul, B. Dou\c cot and R. Mosseri for
fruitful discussions.
\end{acknowledgments}

\appendix
\section{Fourier Transform of Precious Mean Potentials}
\label{ap:precious}

We consider a periodic one-dimensional chain decorated by a diagonal
potential $W$ whose amplitudes
on each site can take two discrete values $W_A=+\lambda/2$ or
$W_B=-\lambda/2$ according to the substitution rule~:
\begin{equation} A \rightarrow A^{k}B,\hspace{2.ex} B\rightarrow A
\mbox{.}
\label{sub.rule}
\end{equation}
To compute the Fourier transform of $W$, it is convenient to consider
the $l^{th}$ order
approximant of the potential obtained by iterating $(l-1)$ times the
rule (\ref{sub.rule}).
We have represented below the first approximants  of the Fibonacci
sequence ($k$=1)~:
\begin{equation}
\begin{array}{ll} l=1 & B \\ l=2 & A \\ l=3 &AB \\ l=4 &ABA\\ l=5
&ABAAB\\ l=6 &ABAABABA
\mbox{.}
\end{array}
\end{equation}
For the $l^{th}$ order approximant, $W$ is thus a periodic potential
with an elementary period
$n_l=F_l$ containing $s_l=G_l$ elements $W_A$ and $p_l=G_{l-1}$ elements
$W_B$ where $(F_{n})_{n\in
{\bf N^*}}$ and $(G_{n})_{n\in {\bf N}}$ are the precious mean
sequences defined by~:
\begin{eqnarray} &&F_1=F_2=1\:;\: G_0=1, \:\: G_1=0\\ &&F_{n+1}=k\,
F_{n}+F_{n-1}\:, \:\:\:\:
\forall n >1 \\ &&G_{n+1}=k\, G_{n}+G_{n-1}\:, \:\: \forall n >0
\mbox{.}
\end{eqnarray}
In the quasiperiodic limit ($l\rightarrow \infty$), the ratio
$s_l/p_l$ converges toward the Pisot
solution\footnote{A number is said to be a Pisot number if it is
solution of a polynomial
equation  with integer coefficients  such that its modulus is bigger
than one and all other
solutions moduli are smaller than one.} $\tau={({k+\sqrt{k^2+4}}) /
2}$ of the equation
$x^2=k\, x+1$. For $k=1,2,3,4$, $\tau$ is known as the golden,
silver,  bronze and chocolate mean respectively.
Since $\tau$ is always irrationnal $(\forall k \in {\bf N^*})$, the
length of the period becomes
infinite, and the sequence is quasiperiodic.

\begin{figure}
\centerline{\includegraphics[width=60mm]{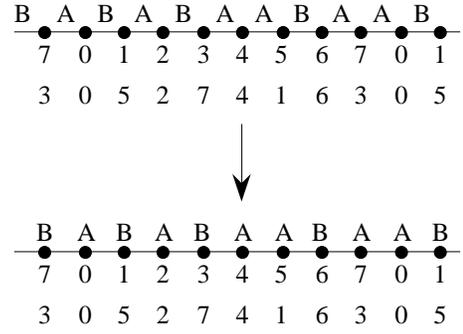}}
\caption{\label{Fibodeco} A possible correspondence between the potential and
the length applied to the
$6^{th}$ approximant of the Fibonacci sequence ($n_6=8$, $s_6=5$, et
$n'_6=5$). The upper line gives
the canonical numbering of the sites and the lower line, the conumbering.}
\end{figure}

The precious mean sequences can also be built by the Cut and Project
algorithm starting from the
usual ${\bf Z}^2$ lattice and choosing, for the cut slope,
$\alpha_l=s_l/p_l$.  In this case, one
obtains a periodic structure with two types of lengths $L_A$ and
$L_B$ distributed according to (\ref{sub.rule}) but with a different
origin than that given by
(\ref{sub.rule}). To establish a correspondence between the potential
and the structure, one can,
for example, affect  $W_A$ (resp. $W_B$) to a site if the adjacent
left segment of this site has
length $L_A$ (resp. $L_B$) as displayed in Fig.~\ref{Fibodeco}.

For this type of tiling (codimension 1), it is possible to label each
site according to
their local environment and thus to classify them with respect to
their on-site
potential\cite{mosseri_conumbering,mosseri_conumbering_proceedings,vid 
al_rauzy}.  This
procedure, known as the conumbering scheme, allows to easily
compute the Fourier
transform of the potential, since it is simply expressed in terms of
the so-called
generating vector.

To achieve this conumering in our case, we introduce, following
Ref.\cite{vidal_rauzy},
the parallel space vector in the ${\bf Z}^2$ lattice for the $l^{th}$
order approximant
${\bf A}_{l}^{\parallel}=(G_l,G_{l-1})$,
and we look for the generating vector ${\bf g}_{l}$ determined by the
condition $\det({\bf
A}_{l}^{\parallel},{\bf g}_{l})=1$. Since, the
precious mean sequences verify the relation\footnote{This identity
can be simply derived by recursion.}~:
\begin{equation} G_{l} G_{l-2}-G_{l-1}^2=(-1)^{l}
\mbox{,}
\end{equation}
the generating vector is straightforwardly given by~:
\begin{equation}
{\bf g}_{l}=(-1)^l(G_{l-1},G_{l-2})
\mbox{.}
\end{equation}

After the projection step, each site can then be indexed by its conumber
$j$ defined by $r_j=j n'_l \: [n_l]$ for $j=[0,n-1]$ where $r_j$ 
denotes the canonical
indexing (see Fig.~\ref{Fibodeco}), where $n'_l=(-1)^l
(G_{l-1}+G_{l-2})$ and where
$n_l= G_{l}+G_{l-1}$. These two numbers  ($n_l$ and $n'_l$) are in
fact the lengths of
the vectors ${\bf A}_{l}^{\parallel}$ and ${\bf g}_{l}$ respectively
measured in the
${\bf Z}^2$ lattice unit. Thus, all the sites whose conumber $j\in [0,s_l-1]$
(resp. $j \in [s_l,n_l-1]$) have a potential $W_A$  (resp. $W_B$).

As a result, the Fourier transform of the potential is simply given by~:
%
%
\begin{eqnarray}
\hat W\left(q={2\pi m\over n}\right)&=&{1 \over n} \sum_{j=0}^{n-1}
W_j e^{i q r_j} \nonumber\\
&=&{1 \over n} \left( W_A \sum_{j=0}^{s-1}  e^{i q j n' [n]}+  W_B
\sum_{j=s}^{n-1}
e^{i q j n' [n]}
\right) \nonumber\\
&=&{\lambda \, e^ {i{\pi m  n' (s-1)
\over n}} \sin
\left({\pi m n' s \over n}\right)
\over n \sin \left({\pi m n'\over n}\right)}
\mbox{,}
\label{TF}
\end{eqnarray}
%
%
for integer values of $m \in[1,n-1]$ and~:
%
%
\begin{equation}
\hat W(0)= \lambda
(s-p)/2n
\mbox{.}
\end{equation}
%
%
Note that in (\ref{TF}), we have omitted the index
$l$ for clarity.

\section{Bosonization of Spinless Fermions}\label{ap:bosonization}

Let us first consider free fermions, {\it i.~e.} with $V=W=0$. The
kinetic part of the Hamiltonian (\ref{eq:hamiltonian}) is
easily diagonalized {\it via} Fourier transform~:
%
%
\begin{equation}
H_t=\sum_{k} \varepsilon (k) c^\dagger_{k}\,c_{k}
\mbox{,}
\label{hamilkinetic}
\end{equation}
%
%
where $c^\dagger_{k}=1/\sqrt{L} \: \sum_j e^{i k j}c^\dagger_{j}$ and
$\varepsilon (k)=-2 t \cos k$.

If one is interested in the low-energy properties of the system, the
only relevant states are those
standing around the Fermi points $\pm k_F$. One can thus
linearize the dispersion
relation around these points and obtain an effective Hamiltonian~:
%
%
\begin{equation}
H_t=v_F \sum_k (k-k_F) c^\dagger_{R, k}\,c_{R, k}- (k+k_F)
c^\dagger_{L, k}\,c_{L, k}
\mbox{,}
\label{hamil_TL}
\end{equation}
%
%
where $v_F=2 t \sin k_F$ is the Fermi velocity and where we have
introduced the right $R$
(respectively the left $L$)  movers fermions with momentum close to
$+k_F$ (respectively to $-k_F$).

We now introduce the  fermions fields~:
%
%
\begin{eqnarray}
\psi_{\nu}(x)&=&{1 \over \sqrt{L}} \sum_k e^{ikx} c_{\nu, k}
\mbox{,}
\end{eqnarray}
%
%
where $\nu=R$ or $L$, so that the Hamiltonian (\ref{hamil_TL}) writes
in the continuum limit~:
%
%
\begin{equation} H_t=- i v_F \int  dx\  \psi^\dagger_R(x)
\partial_x\psi_R(x)-
\psi^\dagger_L(x)\partial_x\psi_L(x)
\label{Dirac}
\mbox{.}
\end{equation}
%
%
We also introduce the  right and left Fourier components of the
fermions density operators~:
%
%
\begin{eqnarray}
\rho_{\nu}(q)&=& \sum_k c^\dagger_{\nu, k+q} c_{\nu, k}
\mbox{,}
\end{eqnarray}
%
%
which satisfy bosonic commutation relations~:
%
%
\begin{eqnarray}
\label{com1}
\left[\rho_R(-q),\rho_R(q')\right]&=&  {q L \over 2 \pi} \delta_{q
q'} \mbox{,}\\
\label{com2}
\left[\rho_L(-q),\rho_L(q')\right]&=& -{q L \over 2 \pi} \delta_{q
q'} \mbox{,}\\
\left[\rho_R(q),\rho_L(q')\right]&=& 0
\mbox{.}
\end{eqnarray}
%
%

The commutation relations of these operators with the Hamiltonian $H_t$~:
%
%
\begin{eqnarray}
\left[H_t,\rho_R(q)\right]&=&  v_F\: q \: \rho_R(q) \mbox{,}\\
\left[H_t,\rho_L(q)\right]&=&- v_F\: q \: \rho_L(q)
\mbox{,}
\end{eqnarray}
%
%
explicitely show that $\rho_{\nu}(q)$ generate eigenstates of
$H_t$ with the
energy $v_F q$. This allows to write the kinetic energy as a
bilinear operator in the bosonic
fields~:
%
%
\begin{equation} H_t={\pi v_F\over L}\sum_{q\ne 0}
\left[\rho_R(q)\rho_R(-q)
+\rho_L(q)\rho_L(-q) \right]
\mbox{,}
\end{equation}
%
%
or in the real space~:
%
%
\begin{equation}
H_t=\pi v_F \int dx \left[\rho_R(x)^2 +\rho_L(x)^2 \right]
\mbox{,}
\end{equation}
%
%
where $\rho_{\nu}(x)=\psi^{\dagger}_{\nu}(x)\psi_{\nu}(x)$ for $\nu=R$ or $L$.

We now introduce the fields $\phi$ and $\theta$~:
%
%
\begin{eqnarray}
\phi(x)&=&{- i \pi \over L} \sum_{q\neq 0}{ (\rho_R(q) + \rho_L(q))
\over q} e^{-iqx}
\mbox{,}\\
\theta(x)&=&{ i \pi \over L}\sum_{q\neq 0}{ (\rho_R(q) - \rho_L(q))
\over q} e^{-iqx}
\mbox{,}
\end{eqnarray}
%
%
together with their conjugate momenta~:
%
%
\begin{eqnarray}
\Pi_{\phi}(x)&=&\displaystyle{{1\over \pi}} \partial_x \theta(x) \mbox{,}\\
&=&\displaystyle{{1 \over L}} \sum_{q\neq 0} \left( \rho_R(q) -
\rho_L(q) \right) e^{-iqx}
\mbox{,}\\
&=& \rho_R(x)- \rho_L(x) \mbox{,}\\
\Pi_{\theta}(x) &=& \displaystyle{{1\over \pi}} \partial_x \phi(x) \mbox{,}\\
&=&\displaystyle{{-1 \over L}}\sum_{q\neq 0} \left( \rho_R(q) +
\rho_L(q) \right) e^{-iqx}
\mbox{,}\\
&=& - \rho_R(x)- \rho_L(x)
\mbox{,}
\end{eqnarray}
%
%
which obey the canonical commutation rules~:
%
%
\begin{eqnarray}
\left[ \phi(x), \Pi_\phi(x) \right]&=& i \delta(x-x') \mbox{,}\\
\left[ \theta(x), \Pi_\theta(x) \right] &=& i \delta(x-x')
\mbox{.}
\end{eqnarray}
%
%
This finally allows to rewrite the kinetic energy~:
%
%
\begin{equation}
H_t={v_F\over 2\pi} \int dx \left[(\pi\Pi)^2
+(\partial_x \phi)^2 \right]
\mbox{,}
\label{freehamil}
\end{equation}
%
%
with $\Pi=\Pi_{\phi}$, which is the Hamiltonian of a one-dimensional
elastic string. The fermions operators, and thus, all physical
quantites (charge density
wave, Cooper pairs density,...), can be easily expressed in terms of
the $\phi$ and $\Pi$
fields. The correspondence between the two sets of operators  are
given by the following
relations~:
%
%
\begin{eqnarray}
\psi_R^\dagger(x)&=&{1 \over \sqrt{2 \pi \alpha}} U^\dagger_R e^{i(\phi(x)-
\theta(x))}\\
\psi_L^\dagger(x)&=&{1 \over \sqrt{2 \pi \alpha}} U^\dagger_L e^{-i(\phi(x) +
\theta(x) )}
\mbox{,}
\end{eqnarray}
%
%
$\alpha$ being an ultraviolet cut-off and  $\theta(x)$ being defined by~:
$\theta(x)=\pi
\int_{-\infty}^x\ \Pi(x') dx'$. The operators $U_R$ and $U_L$  are
anticommuting operators which
ensure the correct commutation relations of the fermions operators.
Note that these operators give
essentially vanishing contributions in the thermodynamical limit and
can be safely ignored, at
least in the case of single chain systems.

The great advantage of this bosonic formulation is that the interaction term
$H_V$ can be almost trivially taken into account. Let us consider,
for instance, a two-body interaction~:
%
%
\begin{equation}
H_{V}={1\over 2}\int  dx \int dx' \rho(x)\ V(x-x') \:\rho(x')
\mbox{,}
\label{interaction}
\end{equation}
%
%
\\
where $\rho(x)$ is the fermion density operator on site $x$.

If one is only interested in the low-energy physics, the fermion
field operator $\psi(x)$ writes~:
%
%
\begin{equation}
\psi(x)=e^{i k_F x} \psi_R(x) + e^{-i k_F x} \psi_L(x)
\mbox{,}
\end{equation}
%
%
which correctly describes the low-energy behavior of the
particle-hole and particle-particle density
state. It follows that the density operator expresses as~:
%
%
\begin{widetext}
\begin{eqnarray}
\rho(x)&=& \psi^\dagger(x) \psi(x)  \mbox{,}\\
            &=& \psi^\dagger_R(x) \psi_R(x) + \psi^\dagger_L(x) \psi_L(x) +
                 e^{-2ik_F x} \psi^\dagger_R(x) \psi_L(x) +
                 e^{2ik_F x}\psi^\dagger_L(x) \psi_R(x) \mbox{,}\\
            &=& \rho_R(x) + \rho_L(x) +e^{-2ik_F x} \psi^\dagger_R(x) 
\psi_L(x) +
                                       e^{2ik_F x}\psi^\dagger_L(x) \psi_R(x)
\label{densite} \mbox{,} \\
            &=& -{1 \over \pi} \partial_x \phi(x) + {1 \over \pi \alpha}
\cos(2\phi(x)-2k_F x)
\mbox{,}
\end{eqnarray}
\end{widetext}
%
%
and~:
%
%
\begin{widetext}
\begin{eqnarray}
\rho(x) \rho(x')&=& \rho_R(x) \rho_R(x') + \rho_L(x) \rho_L(x') +
\rho_R(x) \rho_L(x')
                          + \rho_L(x) \rho_R(x') + \nonumber\\
                        & & e^{-i2k_F(x-x')} \:\psi^\dagger_R(x) \psi_L(x)
                            \psi^\dagger_L(x') \psi_R(x') +  \\
                        & & e^{i2k_F(x-x')} \: \psi^\dagger_L(x) \psi_R(x)
                            \psi^\dagger_R(x') \psi_L(x') + \nonumber\\
                        & & \mbox{oscillating terms} \nonumber
\label{densite-densite}
\mbox{.}
\end{eqnarray}
\end{widetext}
%
%

In the case of short-range interactions the remaining oscillating
terms are supposed to
give negligible contributions and we keep only terms behaving like
$e^{-i 2 k_F (x-x')}$.

If one consider local interactions, one can write the interaction term
in the general form~:
\begin{eqnarray}
H_{V}=
\int dx  \: &&{g_4 \over 2} \bigg[ \rho_R(x) \rho_R(x) + \rho_L(x)
\rho_L(x) \bigg] + \nonumber\\
&&
g_2 \:\rho_R(x) \rho_L(x)
\label{hamil_geol}
\mbox{,}
\end{eqnarray}
with the standard notations of $g-ology$. Gathering (\ref{freehamil})
and (\ref{hamil_geol})
and  the Hamiltonian
$H_0(t,V)$ then becomes~:
\begin{equation}
H_0(t,V)={1 \over 2 \pi} \int dx  \: (uK) (\pi \Pi)^2 +
\left({u \over K}\right)
(\partial_x \phi)^2
\mbox{,}
\label{ham_boso}
\end{equation}
where we have introduced the parameters  $u$ and $K$ (Luttinger
parameter) given by~:
\begin{eqnarray}
u&=&\sqrt{\left(v_F+{g_4 \over 2\pi}\right)^2- \left({g_2 \over
2\pi}\right)^2} \mbox{,} \label{eq:pertlut2} \\
K&=&\sqrt{{2\pi v_F+ g_4 - g_2 \over 2\pi v_F+ g_4 + g_2}} \label{eq:pertlut}
\mbox{.}
\end{eqnarray}

The Hamiltonian (\ref{ham_boso}) can still be interpreted as the
Hamiltonian of an elastic
string with effective parameters $u$ and $K$.

For the model considered here, the interaction term writes~:
%
%
\begin{equation}
V \sum_i n_i n_{i+1}\: \rightarrow
\lim_{a\rightarrow 0} Va \int dx\: \rho(x)\, \rho(x+a)
\mbox{.}
\label{eq:tv}
\end{equation}
%
%
This expression can also be written as (\ref{hamil_geol}) for $V\ll t$ with~:
$g_4=g_2=2 V(1-\cos(2k_F a))$.
Note that, in the non interacting case one has $K=1$ and
$u=v_F=2t\sin(k_F a)$. For
repulsive interactions ($V>0$), $K<1$ while the attractive case ($V<0$)
leads to $K>1$. A remarkable fact is that the representation
(\ref{ham_boso}) is, in fact,
completely general\cite{haldane_xxzchain,haldane_bosonisation} and
gives the correct
low-energy description of the
system, even when the interactions are strong provided the correct
$u$ and $K$ parameters are used.\\\\\\\

For example for $k_F=\pi/2$ one has for the interaction (\ref{eq:tv})~:
\begin{eqnarray}
\frac{V}{2t} &=& - \cos\left( {\pi\over 2K} \right) \\
u & =& \frac{2 K t}{2K-1}
\sin\left(\pi\left(1-\frac1{2K}\right)\right) \label{eq:lutcalc}
\mbox{.}
\end{eqnarray}
Concerning the bosonization of $H_W$, one straightforwardly obtains, in
the continuum limit~:
\begin{widetext}
\begin{equation} H_{W}= \int  dx \: W(x) \, \rho(x)= \int dx \: W(x) \,
\left(-{1 \over \pi} \partial_x
\phi(x) + {1 \over \pi \alpha} \cos(2\phi(x)-2k_F x) \right)
\mbox{.}
\label{defhw}
\end{equation}
\end{widetext}

\section{Renormalization Group Equations}
\label{ap:rgpot}

The aim of this appendix is to derive the RG equations used in
Section \ref{The_Model}, following
the method of Ref.~\onlinecite{giamarchi_spin_flop}. \\
We evaluate the
correlation
function $R(x,\tau,x',\tau')$
perturbatively at second order in power of the coupling constant $g$~:
\begin{widetext}
\begin{eqnarray}
\left\langle T_\tau e^{i  \sqrt{2}\phi(x',\tau')}
e^{-i  \sqrt{2}\phi(x,\tau)}\right\rangle
&=&
I_0 - {g \over u (2\pi \alpha)^2} \: I_1 +
{1 \over 2}{g^2 \over u^2 (2\pi \alpha)^4} \:I_2 +o(g^3)\
\mbox{,}
\end{eqnarray}
\end{widetext}
with~:
\begin{widetext}
\begin{eqnarray}
I_0&=&\left\langle T_\tau e^{i  \sqrt{2}\phi(x',\tau')}
e^{-i  \sqrt{2}\phi(x,\tau)}\right\rangle_0  \label{defI0}\\
I_1&=&\sum_Q {\hat W(Q)} \int d^2{\bf r}_1
\left\langle
T_\tau  e^{i  \sqrt{2}\phi(x',\tau')} e^{-i  \sqrt{2}\phi(x,\tau)}
\left[
e^{i\left[2\phi(x_1,\tau_1) + Q^- x_1 \right]} +
e^{i\left[-2\phi(x_1,\tau_1) + Q^+ x_1 \right]}
\right] \right\rangle_0
\mbox{,}
\label{defI1}
\end{eqnarray}
\end{widetext}
with $Q^\pm=Q \pm 2k_F$ and~:
\begin{widetext}
\begin{eqnarray}
I_2&=&\sum_{Q_1,\,Q_2}
{\hat W(Q_1)}{\hat W(Q_2)} \int d^2{\bf r}_1 \int d^2{\bf r}_2
\left\langle
T_\tau  e^{i \sqrt{2}\phi(x',\tau')} e^{-i  \sqrt{2}\phi(x,\tau)}
\right. \times
\label{defI2}\\
&&
\left[
e^{i\left[2\phi(x_1,\tau_1) + Q^-_1 x_1 \right]} +
e^{i\left[-2\phi(x_1,\tau_1) + Q^+_1 x_1 \right]}
\right]
\left.\left[
e^{i\left[2\phi(x_2,\tau_2) + Q^-_2 x_2 \right]} +
e^{i\left[-2\phi(x_2,\tau_2) + Q^+_2 x_2 \right]}
\right]
\right\rangle_0^c \nonumber
\mbox{,}
\end{eqnarray}
\end{widetext}
where $\langle\rangle_0$ denotes the average performed with respect
to the free action,
{\it i.~e.} with $g=0$. The index
$c$ in Eq.(\ref{defI2})  refers to the connected correlation function
with respect to
the quantities indexed by 1 and 2.

The average in Eqs. (\ref{defI0}) (\ref{defI1}) and (\ref{defI2})  are easily
performed using the relation~:
\begin{widetext}
\begin{equation}
\left\langle  T_\tau e^{ic_1 \phi(x_1,\tau_1) } e^{ic_2 \phi(x_2,\tau_2)}
\dots
e^{ic_n \phi(x_n,\tau_n)}\right\rangle_0 \simeq e^{{K \over 2}
\sum_{i>j} c_i c_j
\ln {|{\bf r}_i-{\bf r}_j| \over \alpha}}
\label{correlateur}
\mbox{,}
\end{equation}
\end{widetext}
when $|{\bf r}_i-{\bf r}_j| \gg \alpha$. Note also that the expression
(\ref{correlateur}) vanishes if the ``neutrality'' condition,  $\sum_{i}
c_i=0$, is not satisfied. One thus has~:
\begin{eqnarray}
I_0&=&\left\langle T_\tau e^{i  \sqrt{2}\phi(x',\tau')}
e^{-i  \sqrt{2}\phi(x,\tau)}\right\rangle_0\\
\nonumber\\
&=&e^{i{\sqrt{2} K \over u} \int_{x'}^x W(y) dy}
\left\langle T_\tau e^{i  \sqrt{2}\tilde\phi(x',\tau')}
e^{-i  \sqrt{2}\tilde\phi(x,\tau)}\right\rangle_0\\
\nonumber\\
&=& e^{i{\sqrt{2} K \over u} \int_{x'}^x W(y) dy}\: e^{-F({\bf r -r'})}
\mbox{,}
\end{eqnarray}
with $F({\bf r -r'})= K \ln |{\bf r -r'}|/\alpha$.

Note that $I_1$ vanishes since it does not respect the neutrality
condition. It follows that the first non trivial contribution of the
potential is provided by the term $I_2$. Keeping in (\ref{defI2})
the terms satisfying the neutrality condition and performing the average
over the fields one obtains~:
\begin{widetext}
\begin{eqnarray}
I_2&=&
e^{i{\sqrt{2} K \over u} \int_{x'}^x W(y) dy} \: e^{-F({\bf r -r'})} \times
\label{I2} \\
&& \sum_{Q_1,\,Q_2}{\hat W(Q_1)}{\hat W(Q_2)} \int d^2{\bf r}_1 \int
d^2{\bf r}_2
\:e^{-2 \, F({\bf r}_1 - {\bf r}_2)}
e^{i{2 K \over u} \int_{x_2}^{x_1} W(y) dy} \times \nonumber \\
&&
\bigg[
e^{-\sqrt{2} \, F({\bf r} - {\bf r}_1) + \sqrt{2} \, F({\bf r} - {\bf r}_2)
+   \sqrt{2} \, F({\bf r'} - {\bf r}_1)- \sqrt{2} \, F({\bf r'} - {\bf r}_2)}
-1 \bigg]
\bigg[
e^{i \left(Q^-_1 x_1+ Q^+_2 x_2\right)} +
e^{i \left(Q^+_1 x_2+ Q^-_2 x_1\right)}\bigg]
\nonumber
\mbox{.}
\end{eqnarray}
\end{widetext}

At this order the potential enters only through a pure phase and does not
drastically affects the long distance behavior. Anyway,  a choice of
a vanishing average value for the potential renders this effect
negligible.

The integration over ${\bf r}_1$ et ${\bf r}_2$ is performed with
the following variable changes~:
\begin{equation}
{\bf r}_1={\bf R} + {\delta{\bf r} \over 2} \:, \:\:
{\bf r}_2={\bf R} - {\delta{\bf r} \over 2}
\mbox{,}
\end{equation}
since, due to the factor $\:e^{-2 \, F({\bf r}_1 - {\bf r}_2)}$ in
Eq.(\ref{I2}) the non vanishing  contributions are given by ``points''
separated by small values of $\delta{\bf r}={\bf r}_1-{\bf r}_2$.
If the potential is translation invariant, the term $e^{i{2 K \over u}
\int_{x_2}^{x_1} W(y) dy}$ does only depends on $x_1-x_2$ and is thus
of order one for
small $\delta{\bf r}$. Of course, in the quasiperiodic case, this is
no longer true.
Nevertheless, if one is only interested by the intrinsic properties
of the potential,
it is reasonable to perform an average over all the possible choice of the
sequence origin. Then, we can also consider that $e^{i{2 K \over u}
\int_{x_2}^{x_1} W(y) dy}$ only depends on $x_1-x_2$ and replace it by 1.

At leading  order in $\delta{\bf r}$, one has~:
\begin{widetext}
\begin{equation}
F({\bf r} - {\bf r}_1) -   F({\bf r} - {\bf r}_2) -
F({\bf r'} - {\bf r}_1) +  F({\bf r'} - {\bf r}_2)
=
\delta{\bf r}. \nabla_{\bf R}
\left[ F({\bf r}'-{\bf R})-F({\bf r}-{\bf R})\right]+o(\delta{\bf r})^3
\mbox{.}
\end{equation}
\end{widetext}

On the other hand, with ${\bf R}=(X,Y)$ et
$\delta{\bf r}=(\delta x, \delta y)$ one has~:
\begin{equation}
Q^-_1 x_1+ Q^+_2 x_2 = \left( Q_1 + Q_2 \right) X +
\left( Q^-_1 - Q^+_2 \right) {\delta x \over 2}
\mbox{,}
\end{equation}
and~:
\begin{equation}
Q^+_1 x_2+ Q^-_2 x_1 = \left( Q_1 + Q_2 \right) X +
\left( Q^-_2 - Q^+_1 \right) {\delta x \over 2}
\mbox{,}
\end{equation}
which implies that the integration over the variable $X$ selects the
contributions $Q_1=-Q_2=Q$ in the sum over $Q_1$ and $Q_2$. Also, taking
the fact that the potential $W$ can be taken as real one has
$\hat{W}(Q)=\hat{W}^*(-Q)$
and~:
\begin{widetext}
\begin{eqnarray}
I_2  &=& e^{i{\sqrt{2} K \over u} \int_{x'}^x W(y) dy} \: e^{-F({\bf r -r'})}
\:\sum_Q \left| \hat W(Q) \right|^2
\int d^2{\bf R} \int d^2\delta {\bf r} \:\: e^{-2 \, F(\delta {\bf r})}
\bigg[
e^{i  \left(Q^- \delta x \right)} +
e^{-i \left(Q^+ \delta x \right)}
\bigg] \times
      \\
&&
\bigg(
\delta{\bf r}. \nabla_{\bf R}
\left[ F({\bf r}'-{\bf R})-F({\bf r}-{\bf R})\right]
\bigg)^2
\nonumber
\mbox{.}
\label{I}
\end{eqnarray}
\end{widetext}
Let us now define~: $\zeta=F({\bf r}'-{\bf R})-F({\bf r}-{\bf R})$. One has~:
\begin{equation}
(\delta{\bf r}. \nabla_{\bf R} \: \zeta)^2
= (\delta x \: \partial_{X} \: \zeta)^2+ (\delta y \:\partial_{Y} \:\zeta)^2 +
2 \delta x \: \delta y \: \partial_{X} \: \zeta \: \partial_{Y} \: \zeta
\mbox{.}
\end{equation}
By parity only the two first terms would survive to the integral over
$\delta y$ (or $\delta x$). On the other  hand, the integral over
$\bf R$ is easily
performed and leads to two kind of terms~:
\begin{equation}
\int d^2{\bf R} \:\zeta \left(\partial^2_{X}+\partial^2_{Y} \right) \zeta =
-4 \pi K^2 \ln {|{\bf r -r'}| \over \alpha}
\mbox{,}
\end{equation}
and~:
\begin{equation}
\int d{\bf R} \:\zeta \left(\partial^2_{X}-\partial^2_{Y} \right) 
\zeta =- 2 \pi
\cos 2 \theta_{\bf r-r'}
\label{thetarenorm}
\mbox{,}
\end{equation}
where $\theta_{\bf r-r'}$ is the angle between the vector $(x,\tau)$ and
the $x$ axis. The occurence of the second term comes from the fact that
the free correlation function $F$ writes, in fact, as~:
\begin{equation}
F({\bf r -r'})=K \ln {|{\bf r -r'}|\over \alpha}+ d \cos (2 \theta_{\bf r-r'})
\mbox{,}
\end{equation}
where $d$ parametrizes the anisotropy between the space and time
direction. One has $d=0$ in the original Hamiltonian but, as seen in
Eq.(\ref{thetarenorm}), this anisotropy is generated by
renormalization. Ultimately, this is equivalent to a
renormalizaton of the Fermi velocity $u$. However, since the anisotropy
itself is of order $g^2$,  the correction to $u$ can be neglected in this
calculation.

Gathering the preceeding results one thus finds~:
\begin{widetext}
\begin{eqnarray}
I_2&=&
e^{i{\sqrt{2} K \over u} \int_{x'}^x W(y) dy} \: e^{-F({\bf r -r'})}
\: 2\pi K^2 \:\sum_Q \left| \hat W(Q) \right|^2
\ln {|{\bf r}-{\bf r}'| \over \alpha} \times \\
&&
\int d \delta {\bf r} \:
\delta^2{\bf r}\: e^{-2 \, F(\delta {\bf r})}
\bigg[
e^{i  \left(Q^- \delta x \right)} +
e^{-i \left(Q^+ \delta x \right)}
\bigg]
\nonumber
\mbox{,}
\end{eqnarray}
\end{widetext}
which also reads~:
\begin{widetext}
\begin{eqnarray}
I_2&=&
e^{i{\sqrt{2} K \over u} \int_{x'}^x W(y) dy} \: e^{-F({\bf r -r'})}
\: 2\pi K^2 \:\sum_Q \left| \hat W(Q) \right|^2
\ln {|{\bf r}-{\bf r}'| \over \alpha} \times  \\
&&
2\pi \alpha^3 \int_\alpha^{+\infty}  \hspace{-2ex}
d \delta r \:
\left({\delta r \over \alpha}\right)^{3-2K}  \hspace{-1ex}
\bigg[ \mbox{J}_0(Q^+ \delta r) + \mbox{J}_0(Q^- \delta r) \bigg]
\nonumber
\mbox{,}
\end{eqnarray}
\end{widetext}
where $\mbox{J}_0$ is a Bessel function.

Finally the correlation function $R(x,\tau,x',\tau')$ writes~:
\begin{widetext}
\begin{eqnarray}
\left\langle T_\tau e^{i  \sqrt{2}\phi(x',\tau')}
e^{-i  \sqrt{2}\phi(x,\tau)}\right\rangle_g
&=&
e^{i{\sqrt{2} K \over u} \int_{x'}^x W(y) dy} \: e^{-F({\bf r -r'})}\left\{1+
{g^2 K^2 \over 8\pi^2 u^2}
\ln {|{\bf r}-{\bf r}'| \over \alpha}
\sum_Q \left| \hat W(Q) \right|^2
\right. \times \\
\nonumber \\
&&
\left.
\int_\alpha^{+\infty} {d \delta r \over \alpha} \:
\left({\delta r \over \alpha}\right)^{3-2K}
\bigg[ \mbox{J}_0(Q^- \delta r) + \mbox{J}_0(Q^+ \delta r) \bigg]
\right\}
\nonumber
\label{final_fonction}
\mbox{.}
\end{eqnarray}
\end{widetext}

By re-exponentiation one has~:
\begin{equation}
R(x,\tau,x',\tau')=\left({\alpha \over |{\bf r -r'}|}\right)^{K_{eff}}
\mbox{,}
\end{equation}
with~:
\begin{widetext}
\begin{equation}
K_{eff}=K-{K^2 \over 2}
\sum_Q y_Q^2
\int_\alpha^{+\infty} {d \delta r \over \alpha} \:
\left({\delta r \over \alpha}\right)^{3-2K}
\bigg[ \mbox{J}_0(Q^+ \delta r) + \mbox{J}_0(Q^- \delta r) \bigg]
\mbox{,}
\label{Keff}
\end{equation}
\end{widetext}
with $y_Q=\lambda \alpha \hat{W}(Q)/u$.

To derive the RG equation, one has to consider an
infinitesimal variation of the  running cut-off $\alpha(l)$ to $\alpha(l)
e^{dl}$ in Eq.(\ref{Keff}). This leads to~:
\begin{eqnarray} {dK\over dl}&=&-K^2 \,\Xi(l)\\ {dy_Q\over dl}&=&(2-K) \, y_Q
\label{RG2}
\mbox {,}
\end{eqnarray}
with~:
\begin{equation}
\Xi(l)= {1\over 2} \sum_Q y_Q^2
\left[\mbox{J}_0(Q^+ \alpha(l)) + \mbox{J}_0(Q^- \alpha(l)) \right]
\mbox{.}
\label{Gl2}
\end{equation}
Here one has to note that the use of a sharp cut-off in real space
leads to the occurrence of
Bessel function ${\mbox{J}}_0$ . This choice is in fact not
satisfying since it does not ensure
the convergence of the sum  (\ref{Gl2}). This is the reason why we
have considered more general
cut-off procedures for which ${\mbox{J}}_0$ is replaced by faster
decreasing functions which
typically satisfy (\ref{eq:cut-offbis}-\ref{eq:cut-off}).

\section{Memory function}
\label{ap:memoire}

In a normal metal (finite conductivity at $\omega=0$) the Kubo formula~:
\begin{equation} \label{eq:kuboap}
\sigma(\omega) = \frac{i}{\omega}\left[
          \frac{2 u K}\pi + \chi(\omega) \right]
\mbox{,}
\end{equation}
implies that $\chi(0) = -2uK/\pi$. Then (\ref{eq:kuboap}) can
be reexpressed in terms of a meromorphic function $M$ through~:
\begin{equation} \label{eq:condmem}
\sigma(\omega) = \frac{i2uK}\pi \frac1{\omega + M(\omega)}
\mbox{,}
\end{equation}
where $M$ is given by~:
\begin{equation}
M(\omega) = \frac{\omega \chi(\omega)}{\chi(0) - \chi(\omega)}
\mbox{.}
\end{equation}
The interest of the function $M$ lies in the fact that, contrarily to the
conductivity itself, one can expect $M$ to have a well behaved
expansion in the
scattering potential $H_W$. Indeed in a simple hydrodynamic
approximation $M(\omega \to 0)$
would simply be the inverse relaxation time $M \sim i/\tau$, leading
to the standard Lorentzian
broadening of the Drude peak. Another way to formulate it is that a
perturbative calculation
of the memory function is close to a perturbative calculation of the
resistivity.
In the lowest order in the scattering potential $H_W$ one gets~:
\begin{equation}
\chi(0) - \chi(\omega) \sim \chi(0)
\mbox{,}
\end{equation}
and~:
\begin{equation}
\omega \chi(\omega) = [\langle F;F\rangle_\omega^0 - \langle F;F
\rangle^0_{\omega=0}]/\omega
\mbox{,}
\end{equation}
where $F$ operator takes into account the fact that the current is not a
conserved quantity $F = [j,H]$ and $\langle F;F \rangle_\omega^0$ stands
for the retarded correlation function of the operator $F$ at
frequency $\omega$. Since $F$ is itself proportional to the scattering
potential, at lowest order the average can be computed with the Hamiltonian
in the absence of scattering potential. This leads to~:
\begin{equation} \label{ap:appro}
M(\omega) = \frac{[\langle F;F\rangle_\omega^0 - \langle F;F
\rangle^0_{\omega=0}]/\omega}{-\chi(0)}
\mbox{,}
\end{equation}
where $\langle \rangle^0$ stands for an average with
(\ref{eq:ham_boso_texte})
only. Since all averages are to be computed with the quadratic
Hamiltonian (\ref{eq:ham_boso_texte})
only, the computation of $M$ is now feasible.


\end{document}